%% file: main.tex
\documentclass{article}

\input{preamble}

\tcbuselibrary{skins}
\usepackage{titlesec}
\usepackage[labelfont=bf, font=small, labelsep=period]{caption}
\usepackage[absolute,overlay]{textpos}
\usepackage{microtype} 

\definecolor{DeepIndigo}{RGB}{10, 20, 100}
\definecolor{ElectricCyan}{RGB}{0, 220, 255}
\definecolor{NeonYellow}{RGB}{255, 230, 0}
\definecolor{VividRed}{RGB}{220, 20, 10}
\definecolor{DarkCrimson}{RGB}{110, 0, 5}
\definecolor{caspianbg}{RGB}{248, 250, 255} 

\hypersetup{
    colorlinks=true,
    linkcolor=DeepIndigo,
    citecolor=DeepIndigo,
    urlcolor=ElectricCyan
}

\titleformat{\section}
  {\normalfont\Large\bfseries\color{DeepIndigo}}{\thesection}{1em}{}
\titleformat{\subsection}
  {\normalfont\large\bfseries\color{DeepIndigo!80}}{\thesubsection}{1em}{}

\newcommand{\caspianlogo}{%
\begingroup
\bfseries\LARGE
\color{DeepIndigo}C%
\color{DeepIndigo!70!ElectricCyan}A%
\color{ElectricCyan!80!NeonYellow}S%
\color{NeonYellow}P%
\color{NeonYellow!70!VividRed}I%
\color{VividRed}A%
\color{DarkCrimson}N%
\endgroup
}

\begin{document}

\begin{center}
    \vspace*{-1.2cm}
    \begin{minipage}[c]{0.25\linewidth}
        \includegraphics[height=0.85cm]{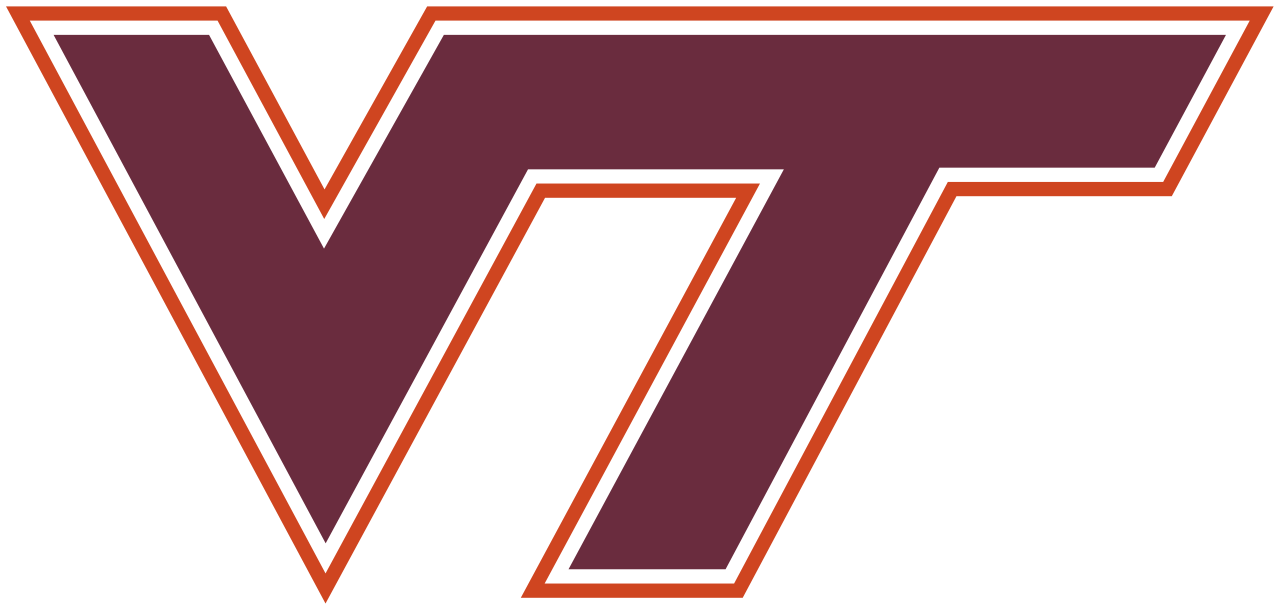}
    \end{minipage}%
    \begin{minipage}[c]{0.5\linewidth}
        \centering
        \footnotesize\textcolor{gray!70}{Preprint \\ \today }
    \end{minipage}%
    \begin{minipage}[c]{0.25\linewidth}
        \hfill 
    \end{minipage}
    
    \vspace{0.25cm}
    {\color{DeepIndigo!20}\rule{\linewidth}{0.6pt}}
    
    \vspace{0.4cm}
\end{center}

\begin{center}
{\LARGE \bfseries
\caspianlogo%
: Online Detection and Attribution of\\
Cascade Attacks in LLM Multi-Agent Systems via\\
Cross-Channel Causal Monitoring
\par}

\vspace{0.25cm}

{\large
Kavana Venkatesh \quad
Jafar Isbarov \quad
Saad Amin \\}
{\large
Murat Kantarcioglu \quad
Jiaming Cui}\\
\vspace{0.05cm}
{\texttt{Virginia Tech, Blacksburg, VA}\\}
{\small
\texttt{\{kavanav, isbarov, saad05, muratk, jiamingcui\}@vt.edu}}\\

\vspace{0.15cm}

{\small \faGithub \quad \href{https://github.com/caspian-detector/caspian}{\textcolor{academicblue}{\texttt{https://github.com/caspian-detector/caspian}}}}

\vspace{0.5cm}
\end{center}

\renewenvironment{abstract}{\noindent\ignorespaces}{\par}

\begin{center}
\begin{tcolorbox}[
    enhanced,
    colback=caspianbg,
    colframe=gray!5,
    arc=0mm,
    outer arc=0mm,
    width=\linewidth,
    left=4mm,
    right=4mm,
    top=3mm,
    bottom=3mm,
    boxrule=0pt,
    borderline west={2.5pt}{0pt}{DeepIndigo!90!ElectricCyan}, 
    before skip=10pt,
    after skip=20pt
]
\input{sec/0_abstract}
\end{tcolorbox}
\end{center}

\vspace{0.1cm}

\input{sec/1_intro}
\input{sec/2_related}
\input{sec/3_background}
\input{sec/4_method}
\input{sec/6_experiments}

\input{sec/7_limitations}

\clearpage

{\small
\bibliographystyle{plain}
\bibliography{main}
}

\clearpage
\input{sec/X_supp}


\end{document}

%% file: preamble.tex
\usepackage[preprint]{neurips_2026}
\usepackage{graphicx}
\usepackage{tikz}
\usepackage[absolute,overlay]{textpos}
\usepackage{tcolorbox}
\usepackage{array}
\usepackage{multirow}
\usepackage{xcolor}
\usepackage{fontawesome5}
\usepackage{hyperref} 
\usepackage{fontawesome5}

\definecolor{academicblue}{RGB}{0, 51, 102} 

\AddToHook{shipout/before}{%
  \pdfpageattr{/Group<</S/Transparency /I true /CS/DeviceRGB>>}%
}

\usepackage[table]{xcolor} 
\usepackage{booktabs}
\usepackage{graphicx}

\definecolor{tabletint}{RGB}{242, 246, 250} 
\definecolor{academicblue}{RGB}{0, 51, 153} 

\usepackage[utf8]{inputenc}
\usepackage[T1]{fontenc}
\usepackage{microtype}

\usepackage{hyperref}
\usepackage{url}

\usepackage{amsfonts}
\usepackage{nicefrac}
\usepackage{siunitx}

\usepackage{booktabs}
\usepackage{multirow}
\usepackage{makecell}
\usepackage{threeparttable}
\usepackage{adjustbox}
\usepackage{tabularx}
\usepackage{graphicx}
\usepackage{array}
\usepackage[table]{xcolor}
\usepackage{colortbl}
\usepackage{soul}
\usepackage{tikz}
\usepackage{amsmath}
\usepackage[table]{xcolor}
\usepackage{amsmath, amsthm, amsfonts}
\newtheorem{definition}{Definition}
\usepackage{empheq}
\usepackage{tcolorbox}
\usepackage{titletoc}
\usepackage{algorithmic}
\usepackage{algorithm}

\newcolumntype{Y}{>{\centering\arraybackslash}X}

\definecolor{lightgray}{gray}{0.95}
\definecolor{headbg}{RGB}{245,247,250}
\definecolor{subheadbg}{RGB}{250,251,253}
\definecolor{groupbg}{RGB}{248,249,251}
\definecolor{nacol}{RGB}{150,150,150}
\definecolor{bestgray}{gray}{0.90}
\definecolor{best}{RGB}{198,239,206}
\definecolor{second}{RGB}{226,239,218}
\definecolor{head}{RGB}{245,246,250}
\definecolor{subhead}{RGB}{250,250,252}
\definecolor{light}{RGB}{248,249,251}

\definecolor{pastelorange}{RGB}{232,150,75}
\definecolor{pastelblue}{RGB}{70,145,210}

\DeclareRobustCommand{\circnum}[1]{%
  \tikz[baseline=(char.base)]{
    \node[
      circle,
      draw=black,
      line width=0.55pt,
      inner sep=0.35pt,
      minimum size=0.85em
    ] (char) {\scriptsize\bfseries #1};
  }%
}

%% file: sec/0_abstract.tex
Cascade attacks in LLM multi-agent systems (MAS) arise when adversarial influence propagates across agents and leads to escalated system-level failures through complex agent interactions. Detecting such cascades is challenging, as their signals are distributed, tightly coupled across interaction channels, and often appear plausibly benign locally but may unfold quickly either within a single turn or gradually across multiple turns. Existing defenses, being largely local and text-centric, fail to capture such cross-channel, temporally coordinated dynamics of cascade propagation. Therefore, we propose \textbf{CASPIAN}, the first framework that provides a unified, cross-channel causal analysis of cascade behavior in LLM-MAS through online monitoring of dynamic influence propagation across agents. CASPIAN models multi-agent interactions using a unified, dynamic causal influence matrix across channels, estimated efficiently via a late-interaction conditional transfer entropy (LI-CTE) formulation, thereby enabling the detection of cascade onset from emergent system-level structure rather than isolated anomalies. It further performs online causal attribution, identifying the origin, bridge, and amplifier agents driving the cascade and reconstructing its principal propagation pathways, capabilities not supported by existing methods. Across diverse multi-agent frameworks and benchmarks, CASPIAN consistently outperforms semantic guardrails, LLM-based judges, and graph-based anomaly detectors in both detection accuracy and early cascade identification while operating with sub-1\% relative overhead latency. These results demonstrate that unified cross-channel causal modeling is essential for reliably detecting and understanding cascade failures in LLM multi-agent systems. 

%% file: sec/1_intro.tex
\section{Introduction}
\label{sec:intro}

As large language models are increasingly deployed within 
collaborative multi-agent systems 
(LLM-MAS)~\cite{wu2024autogen, hong2023metagpt, venkatesh2026physicsagentabm, venkatesh2026crea}, the 
nature of adversarial failures has fundamentally shifted. 
While single-agent systems are vulnerable to isolated 
hallucinations or prompt injection attacks, multi-agent 
architectures are susceptible to \textit{cascade attacks}: coordinated failures in which adversarial or erroneous 
influence injected at one agent propagates, amplifies, and 
self-reinforces across agents and 
turns~\cite{xie2026spark, lee2025prompt}. These failures exploit 
the interaction structure of the system itself: a single 
injected prompt can corrupt shared memory, redirect tool 
outputs, and synchronize agents toward a collectively 
flawed trajectory, producing system-wide failures that are 
difficult to contain once established~\cite{xie2026spark,de2025open}. 
Cascades manifest in two regimes: \textit{abrupt} 
cascades, where strong amplification and synchronization 
produce immediate system-wide coupling within a single 
interaction turn; and \textit{gradual} cascades, where 
locally benign influence accumulates across turns through 
persistent memory reuse, tool artifact propagation, and 
slow semantic drift~\cite{xie2026spark,jiang2026agentlab,
an2026aciarena}.

\begin{figure*}[t]
    \centering
    \includegraphics[width=\textwidth]{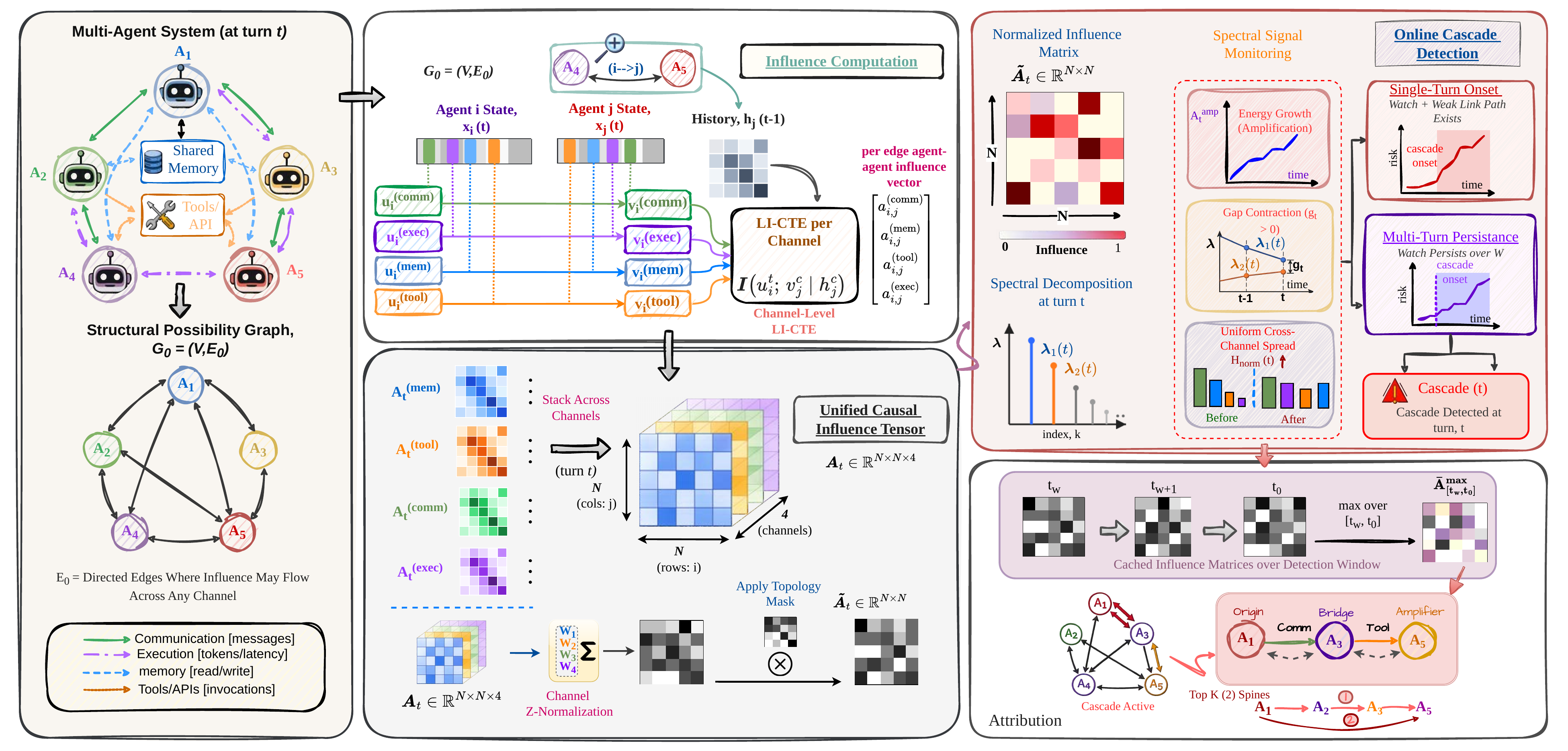}
    \caption{\textbf{Overview of CASPIAN.} CASPIAN constructs a unified cross-channel causal influence tensor from communication, memory, tool, and execution interactions using an efficient LI-CTE-based influence estimation. Spectral monitoring over the evolving topology detects abrupt and persistent cascade formation online, while cached influence dynamics enable recovery of cascade agents, and dominant spines at detection time, with sub-1\% relative overhead latency.}
\label{fig:architecture-overview}
\end{figure*}

The central challenge is that cascade evidence is 
distributed across both interaction channels and turns. 
Because influence may propagate concurrently through 
communication, memory, tool, and execution interactions, 
any single channel provides only a partial view of the 
underlying influence dynamics~\cite{de2025open}. At 
the same time, influence accumulates gradually through 
agent context, so behavior at a given interaction turn may 
still reflect residual effects of injections occurring many 
turns earlier. As a result, agents may appear locally 
benign even as the system collectively transitions toward 
coordinated failure. Detecting cascades therefore requires 
monitoring how influence propagates jointly across agents, 
channels, and time rather than inspecting interactions in 
isolation.

Existing defenses fail to provide this view. Semantic 
guardrails~\cite{chennabasappa2025llamafirewall,
zhang2025jailguard, shi2025promptarmor} and LLM-based 
judges inspect message content per-turn, remaining blind 
to cross-channel propagation structure and temporal 
accumulation. Graph-based 
approaches~\cite{wang2025g,miao2025blindguard,
zhou2026guardian} model agent interactions over 
communication graphs and detect anomalous agents through 
representation-space irregularities, but largely operate on 
a single interaction channel and do not capture how 
influence propagates jointly across memory, tool, and 
execution pathways. Existing attribution methods further 
rely primarily on post-hoc trace analysis over completed 
executions~\cite{zhang2025agent, wang2026flat, zhang2025agentracer}, providing limited 
support for online intervention during active propagation. The shared limitation across these approaches is that they 
reason locally about prompts, agents, or static graphs, 
whereas cascade attacks are fundamentally global 
propagation processes. CASPIAN therefore shifts cascade 
detection from localized semantic inspection to online 
monitoring of propagation dynamics.

The key insight of this work is that cascade onset constitutes a structural reorganization of inter-agent influence propagation in a multi-agent system. Rather than arising from isolated anomalous interactions, cascades emerge through the coupled evolution of four propagation dynamics: growing influence energy (\textit{amplification}), increasing alignment between propagation modes (\textit{synchronization}), spread across interaction channels (\textit{cross-channel propagation}), and sustained reinforcement across turns (\textit{persistence})~\cite{de2025open, xie2026spark, venkatesh2026agent}. Spectral analysis has a well-established role in studying cascade behavior in networked systems~\cite{zocca2021spectral,restrepo2006spectral}, and provides a unified mechanism for tracking these dynamics through the evolving influence topology. Synchronization and persistence manifest as tightening spectral coupling, while amplification and cross-channel propagation reflect increasing cascade reach. In contrast, benign collaboration may temporarily increase interaction intensity without sustained coupling or persistent cross-channel spread~\cite{goltsev2012localization}. Because spectral analysis operates directly on the evolving topology, it requires no supervision or topology-specific calibration.

We present \textbf{CASPIAN}, the \textit{first framework} 
for unified online detection and attribution of cascade 
attacks in LLM-based multi-agent systems through 
cross-channel causal propagation monitoring. Unlike prior 
spectral approaches that operate over static communication 
graphs or pre-defined network structure, CASPIAN models 
the MAS as an evolving cross-channel causal influence 
topology spanning communication, memory, tool, and 
execution interactions. This topology is estimated online 
using a Late-Interaction Conditional Transfer Entropy 
(LI-CTE) formulation~\cite{khattab2020colbert} that 
captures directed, history-aware influence propagation 
with millisecond-scale overhead. CASPIAN then performs 
online spectral monitoring over the evolving topology to 
detect cascade onset through propagation signatures such as 
amplification growth, tightening spectral coupling, phase 
transitions, and cross-channel spread. Together, these 
signals distinguish transient collaborative activity from 
self-reinforcing cascade formation, enabling detection of 
both abrupt single-turn and gradual multi-turn cascades 
without attack templates, benign calibration traces, or 
offline training. Upon detection, CASPIAN further performs 
online attribution directly from cached influence dynamics, 
identifying the cascade origin, amplifier, bridge, and 
dominant propagation spines responsible for system-wide 
spread, enabling timely intervention during active 
propagation. Across diverse multi-agent frameworks and cascade attack 
settings, CASPIAN consistently outperforms semantic 
guardrails, LLM-based judges, and graph-based anomaly 
detectors in both detection accuracy and early cascade 
identification while operating with sub-1\% relative overhead latency.

\begin{itemize}
\item We introduce the \textit{first unified framework} for modeling cross-channel causal influence in LLM multi-agent systems, jointly capturing directed inter-agent influence across communication, memory, tool, and execution channels.
\item We propose an online spectral cascade detection framework that identifies both abrupt single-turn and gradual multi-turn cascades from evolving inter-agent influence dynamics, without attack templates, benign calibration traces, or offline training, with only millisecond-scale overhead.
\item We introduce an online cascade attribution scheme that identifies cascade origins, amplifiers, bridges, and dominant propagation paths directly from cached influence dynamics at detection time, enabling real-time intervention without replay or recomputation.
\end{itemize}

%% file: sec/2_related.tex
\section{Related Work}
\label{sec:related}

\vspace{-6pt}
\subsection{Cascade Failures and Vulnerabilities in LLM-MAS}

LLM-based multi-agent systems are susceptible to cascading failures in which early errors or adversarial signals become self-reinforcing across agents and turns. Prior work identifies cascade amplification, topology sensitivity, and consensus inertia as key drivers of system-wide failure~\cite{xie2026spark,shen2025understanding}. Benchmarks such as TAMAS~\cite{kavathekar2025tamas}, AgentLAB~\cite{jiang2026agentlab}, and ACIArena~\cite{an2026aciarena} expose vulnerabilities from long-horizon prompt injections and cascading agent compromise, while other studies identify memory faults, self-replicating attacks, and cross-channel propagation as major challenges~\cite{zhu2025llm,lee2025prompt,liang2025tipping,de2025open}. Existing approaches largely characterize cascades through attack outcomes, localized message anomalies, or static interaction structure rather than evolving cross-channel propagation during online execution.

\vspace{-6pt}
\subsection{Detection and Defense in LLM-MAS}

Existing defenses against adversarial attacks in LLM-MAS largely fall into three categories. \textit{Semantic guardrails}, including PromptGuard~2~\cite{chennabasappa2025llamafirewall}, JailGuard~\cite{zhang2025jailguard}, and PromptArmor~\cite{shi2025promptarmor}, detect malicious content at the message level, while perplexity-based methods identify token-level anomalies~\cite{radford2019language}. \textit{LLM-based judges}~\cite{gu2024survey} extend inspection to interaction windows or full traces, but remain primarily text-centric. \textit{Graph-based detectors} such as G-Safeguard~\cite{wang2025g}, BlindGuard~\cite{miao2025blindguard}, GUARDIAN~\cite{zhou2026guardian}, and CoopGuard~\cite{EMRA} detect anomalous behaviors through communication topology and representation-space signals. Failure attribution has similarly focused on post-hoc trace analysis~\cite{zhang2025agent,wang2026flat,zhang2025graphtracer,zhang2025agentracer}, while governance frameworks provide oversight mechanisms~\cite{xie2026spark,jia2026mas}. In contrast, CASPIAN models evolving causal influence jointly across communication, memory, tool, and execution channels for online cascade detection and attribution.

\subsection{Causal Influence and Spectral Monitoring}

Prior work has used transfer entropy and related 
information-theoretic measures to estimate directed 
influence in complex 
systems~\cite{schreiber2000measuring,
shahsavari2020estimating,sun2014causation,
lee2024identifying}. Spectral methods have similarly been 
used to analyze propagation and stability in networked 
systems~\cite{chung1997spectral,newman2011structure,
seneta1981nonnegative,horn2012matrix,
hoory2006expander,levin2017markov,
restrepo2006spectral,goltsev2012localization}, while 
path-based models characterize how failures spread through 
network bottlenecks and interaction 
pathways~\cite{watts2002simple,kempe2003maximizing}. 
However, these approaches have not been integrated into an 
online framework for monitoring cross-channel cascade 
formation in LLM-MAS. CASPIAN combines causal influence 
estimation and spectral monitoring to detect and attribute 
cascade propagation across communication, memory, tool, 
and execution interactions during system execution.

%% file: sec/3_background.tex
\section{Background}
\label{sec:background}

\vspace{-6pt}
\subsection{Cascade Failures in LLM Multi-Agent Systems}

LLM-MAS operate through iterative interactions, where agent outputs become inputs to downstream agents across multiple turns, inducing directed dependencies through which influence propagates and reinforces over time~\cite{de2025open,xie2026spark}. A key challenge is the emergence of \textbf{cascade failures}, where erroneous or adversarial signals become progressively self-reinforcing across agents and turns. This behavior is driven by four mechanisms: \circnum{1}~\textbf{Amplification}, where downstream agents reuse corrupted intermediate states~\cite{zhu2025llm,jiang2026agentlab}; \circnum{2}~\textbf{Cross-Channel Propagation}, where influence spreads across communication, memory, and tool interactions~\cite{an2026aciarena,de2025open}; \circnum{3}~\textbf{Persistence}, where signals remain embedded across turns~\cite{xie2026spark}; and \circnum{4}~\textbf{Synchronization}, where agents converge toward coupled reasoning trajectories~\cite{xie2026spark,de2025open}. These dynamics produce both abrupt \textbf{Single-turn Cascades} and gradual \textbf{Multi-turn Cascades}~\cite{an2026aciarena,jiang2026agentlab,zhu2025llm}. Distinguishing transient activity from sustained cascade formation requires analyzing evolving propagation dynamics instead of isolated agent outputs~\cite{watts2002simple,kempe2003maximizing}.

\subsection{Spectral Structure of Directed Influence}
\vspace{-6pt}
A multi-agent system with $N$ agents can be represented by a directed influence matrix $A_t \in \mathbb{R}_{\geq 0}^{N \times N}$ at turn $t$, where $A_t(i,j)$ quantifies the influence of agent $i$ on agent $j$. The spectral structure of $A_t$ characterizes how influence propagates through the system~\cite{chung1997spectral,newman2011structure}. Let $\lambda_1(t) \geq \lambda_2(t)$ denote the two leading ordered propagation spectral values of $A_t$, computed as the largest singular values of the directed influence operator. Here, $\lambda_1(t)$ reflects overall propagation intensity~\cite{seneta1981nonnegative,horn2012matrix}, while $\lambda_2(t)$ captures secondary propagation modes~\cite{chung1997spectral}. Their relationship is governed by the \textit{spectral gap}, $\lambda_1(t)-\lambda_2(t)$: large gaps correspond to well-separated propagation dynamics where perturbations dissipate rapidly, whereas shrinking gaps indicate increasing coupling between propagation modes, enabling influence to persist and reinforce over time~\cite{hoory2006expander,levin2017markov,spielman2007spectral}. Healthy systems may exhibit high activity while maintaining stable propagation structure, whereas cascades are characterized by growing propagation intensity together with tightening spectral coupling~\cite{restrepo2006spectral,goltsev2012localization}. Temporal shifts in this relationship provide a natural signature of cascade onset.



%% file: sec/4_method.tex
\section{Methodology}
\label{sec:method-copy}

\subsection{Problem Setup}

We consider an LLM-based multi-agent system with an agent set 
$\mathcal{V} = \{1,\dots,N\}$ operating over discrete interaction turns 
$t = 1,\dots,T$. At each turn, agents interact through channels 
$\mathcal{C} = \{\texttt{comm}, \texttt{mem}, \texttt{tool}, \texttt{exec}\}$, 
corresponding to inter-agent communication, shared memory access, 
tool-mediated outputs, and execution-level behavior. The system produces an 
online interaction trace up to turn $t$, denoted by $\mathcal{D}_{1:t}$, 
consisting of communication traces, memory operations, tool invocations, and 
execution metadata. We define:

\begin{definition}[Cascade Attack]
A \textbf{cascade attack} is a self-reinforcing propagation process in which 
adversarial or erroneous influence spreads across agents and turns through 
amplification, persistence, and synchronization, either within a single 
interaction channel or across multiple channels over one or more turns.
\end{definition}

Accordingly, we model the system through a unified nonnegative causal influence 
tensor 
$\mathcal{A}_t \in \mathbb{R}_{\ge 0}^{N \times N \times |\mathcal{C}|}$, 
where $\mathcal{A}_t(i,j,c)$ denotes the directed causal influence of agent 
$i$ on agent $j$ through interaction channel $c$ at turn $t$. Larger values 
indicate stronger directed dependence from agent $i$ to agent $j$ through 
channel $c$. The tensor $\mathcal{A}_t$ is latent and must be inferred online 
from the observed trace $\mathcal{D}_{1:t}$. Under this formulation, cascade 
onset corresponds to coordinated structural changes in the evolution of 
$\mathcal{A}_t$.

CASPIAN solves two inference objectives over $\mathcal{A}_t$. 
\textbf{(i) Detection}: infer whether the system has entered a cascade state, 
$\mathrm{Cascade}(t)\in\{0,1\}$, and classify it as either an abrupt 
\textit{Single-turn Cascade} or a gradual \textit{Multi-turn Cascade}. 
\textbf{(ii) Attribution}: upon detection, recover from $\mathcal{A}_t$ 
(i) the \textit{origin}, where abnormal influence first enters the system; 
(ii) the \textit{amplifier}, which most strongly reinforces propagation; 
(iii) the \textit{bridge}, which redistributes influence across the network; 
and (iv) the principal \textit{propagation spines} through which the cascade 
spreads. Algorithm~\ref{alg:caspian} details our method.

\vspace{-6pt}
\subsection{Unified Causal Influence Modeling}
\vspace{-6pt}
We model inter-agent influence through a unified channel-aware causal influence tensor, enabling both global spectral analysis and channel-level reasoning.

Given a MAS with agent set $V=\{0,\dots,N-1\}$ operating over turns $t=1,\dots,T$, we construct a structural possibility graph $G_0=(V,E_0)$ encoding the interaction topology. An edge $(i \rightarrow j)\in E_0$ exists if agent $i$ can influence agent $j$ through at least one channel 
$c \in \mathcal{C}$, where 
$\mathcal{C}=\{\texttt{comm},\texttt{mem},\texttt{tool},\texttt{exec}\}$. For each feasible edge $(i,j)\in E_0$, channel $c$, and turn $t$, CASPIAN estimates directed influence using a temporally-adapted conditional transfer entropy (CTE) formulation. Direct CTE computation over full agent states is noisy and computationally expensive, so we adopt a late-interaction formulation inspired by ColBERT-style decomposition~\cite{khattab2020colbert}, introducing only millisecond-scale overhead:
\begin{equation}
a_{ij}^{(c)}(t)
=
\mathcal{I}\!\left(
\mathbf{u}_{i \to j}^{(c)}(t)\,;\,
\mathbf{v}_{i \to j}^{(c)}(t)
\mid
\mathbf{h}_j^{(c)}(t-1)
\right),
\label{eq:licte}
\end{equation} 
where $\mathbf{u}_{i \to j}^{(c)}(t)$ and 
$\mathbf{v}_{i \to j}^{(c)}(t)$ are compact source and target projections along channel $c$, and $\mathbf{h}_j^{(c)}(t-1)$ is an exponential moving average of the target agent's historical states.

The channel-wise influences are aggregated as

\begin{equation}
\mathbf{a}_{ij}(t)
=
\big[
a_{ij}^{(\texttt{comm})}(t),\;
a_{ij}^{(\texttt{mem})}(t),\;
a_{ij}^{(\texttt{tool})}(t),\;
a_{ij}^{(\texttt{exec})}(t)
\big],
\end{equation}

and stacked to form the causal influence tensor
$\mathcal{A}_t \in \mathbb{R}^{N \times N \times |\mathcal{C}|}$, where $\mathcal{A}_t(i,j,:) = \mathbf{a}_{ij}(t)$. For spectral analysis, the channel matrices are normalized and aggregated into a unified influence matrix $A_t \in \mathbb{R}^{N \times N}$. We then apply degree-aware normalization:

\[
\tilde{A}_t(i,j)=\frac{A_t(i,j)}
{\sqrt{r_i(t)c_j(t)}+\epsilon}
\]

where $r_i(t)$ and $c_j(t)$ denote outgoing and incoming influence strengths.
The tensor $\mathcal{A}_t$ preserves channel-specific influence structure, while
$\tilde{A}_t$ provides a unified representation for cascade detection and
attribution. We use causal influence in an operational, trace-conditioned sense:
source-agent activity is considered influential when it provides directed
predictive information about downstream target behavior beyond the target's own
channel history and the structural possibility graph. See
Appendix~\ref{supp-sec:li-cte} for details.

\vspace{-6pt}
\subsection{Cascade Detection via Spectral Phase Dynamics}
\vspace{-6pt}
After each interaction turn $t$, the unified causal influence tensor 
$\mathcal{A}_t$ is updated using adaptive cumulative LI-CTE estimates over 
agent pairs and interaction channels. CASPIAN then detects cascades 
through spectral transitions in the normalized  matrix $\tilde{A}_t$, characterized by growing influence, increasing 
spectral coupling, cross-channel propagation, and persistence. Detection 
proceeds in two stages: per-turn signal measurement followed by cascade 
classification.

\subsubsection{Per-turn Signal Measurement}

\textbf{Amplification:} Let $\lambda_1(t) \geq \lambda_2(t)$ denote the
two leading ordered propagation spectral values of $\tilde A_t$, computed as
the largest singular values of the directed normalized influence operator. We define the dominant spectral energy 
$E_t = \lambda_1(t) + \lambda_2(t)$
and its turn-over-turn growth as:

\begin{equation}
A_t^{\text{amp}} = \frac{E_t}{E_{t-1} + \epsilon}
\end{equation}

$A_t^{\text{amp}} > 1$ reflects growing influence, but may also arise from benign activity bursts~\cite{restrepo2006spectral}. 

\textbf{Structural Coupling:} The distinguishing factor between transient activity and cascades lies in how influence is organized across propagation modes. Hence, we define spectral coupling ratio $R_t$, normalized spectral gap $g_t$, and gap contraction $\Delta g_t$ as below:

\begin{equation}
R_t = \frac{\lambda_2(t)}{\lambda_1(t) + \epsilon}, 
\qquad
g_t = 1 - R_t,
\qquad
\Delta g_t = g_{t-1} - g_t
\end{equation}

A positive $\Delta g_t$ indicates increasing coupling between the dominant and
secondary propagation modes, reducing separation between pathways and enabling
influence to persist rather than dissipate.

Cascade onset corresponds to a transition in propagation structure rather than gradual drift~\cite{goltsev2012localization}. To capture this, we define:

\begin{equation}
\Phi_t = \frac{|R_t - R_{t-1}|}{R_{t-1} + \epsilon},
\qquad
\textsc{PhaseShift}(t) = \mathbf{1}[\Phi_t > \Delta g_t]
\end{equation}

$\Phi_t$ measures the relative change in spectral coupling across consecutive
turns. A \textbf{Phase Shift} marks entry into a new propagation structure.

\textbf{Cross-Channel Propagation:}
Cascades in LLM-MAS extend across interaction channels, increasing both reach
and persistence~\cite{an2026aciarena,lee2025prompt}. For each channel
$c \in \mathcal{C}$, let $\tilde A_t^{(c)}$ denote the channel-specific
normalized influence matrix, and define its channel energy as
\[
e_c(t) = \lambda_1(\tilde A_t^{(c)}),
\]
where $\lambda_1(\tilde A_t^{(c)})$ denotes the largest ordered propagation
spectral value of the channel-specific directed influence operator. We compute
the normalized channel share as
\[
p_c(t) =
\frac{e_c(t)}
{\sum_{c' \in \mathcal{C}} e_{c'}(t) + \epsilon}
\]

We then measure cross-channel spread using normalized Shannon entropy~\cite{shannon1948}:
\begin{equation}
H^{\mathrm{norm}}_t =
\frac{-\sum_{c \in \mathcal{C}} p_c(t)\log p_c(t)}
{\log |\mathcal{C}|},
\qquad
\textsc{CrossChannel}(t) =
\mathbf{1}[H^{\mathrm{norm}}_t \geq 0.5]
\end{equation}
Higher entropy indicates that propagation energy is distributed across multiple
interaction channels rather than concentrated in a single modality.

\subsubsection{Cascade Classification}

The signals above are combined into a \textsc{Watch} condition that marks the earliest detectable onset of cascade-like dynamics. \textsc{Watch} activates when influence amplification coincides with tightening spectral synchronization, with growth driven by the dominant propagation mode:
\begin{equation}
\textsc{Watch}(t)
=
\mathbf{1}\!\left[
A_t^{\mathrm{amp}}>1
\;\wedge\;
\Delta g_t>0
\;\wedge\;
\lambda_1(t)>\lambda_1(t-1)
\right]
\end{equation}
This requires propagation growth to be driven by the dominant spectral mode 
rather than redistribution across weaker propagation modes.

\noindent \textbf{Weak-link propagation feasibility:} To avoid declaring cascades from isolated high-weight edges, CASPIAN checks
whether the current topology contains a sufficiently strong end-to-end
propagation route. Let $\mathcal{P}(G_0)$ be the set of simple directed paths in
the structural possibility graph with length at most $\mathrm{diam}(G_0)$, and
let $w_t(i,j)=\tilde A_t(i,j)$. We define the widest-path bottleneck score as
\[
B_t =
\max_{\pi \in \mathcal{P}(G_0)}
\min_{(i,j)\in \pi} w_t(i,j)
\]
We compare this bottleneck against the graph's energy-weighted influence scale,
\[
\bar w_t^{\,\mathrm{eng}}
=
\frac{\sum_{(i,j)\in E_0} w_t(i,j)^2}
{\sum_{(i,j)\in E_0} w_t(i,j)+\epsilon}
\]
The weak-link condition is
\begin{equation}
\textsc{WeakLink}(t)
=
\mathbf{1}\!\left[
B_t \geq \bar w_t^{\,\mathrm{eng}}
\right].
\end{equation}
This check requires at least one propagation path whose weakest edge is no weaker than the current graph-level influence scale.

\textit{\textbf{Single-turn cascades:}} A single-turn cascade occurs when \textsc{Watch} coincides with either a phase
shift or cross-channel spread, together with an immediately feasible propagation
route. Let $t_w$ denote the first turn at which \textsc{Watch} activates. For a
single-turn cascade, the confirmation turn coincides with the \textsc{Watch} onset, so
$t_0=t_w$. CASPIAN declares an instant cascade when

\begin{empheq}[box=\fbox]{equation}
\mathrm{\textsc{InstantCascade}}(t_0)
=
\Bigl[
\mathrm{\textsc{Watch}}
\;\wedge\;
(\mathrm{\textsc{PhaseShift}}\vee\mathrm{\textsc{CrossChannel}})
\;\wedge\;
\mathrm{\textsc{WeakLink}}
\Bigr](t_0)
\label{eq:instant_cascade}
\end{empheq}

Here, \textsc{PhaseShift} captures abrupt changes in propagation structure, \textsc{CrossChannel} captures distributed propagation across interaction
modalities, and \textsc{WeakLink} verifies that the topology supports a viable
cascade path.

\textit{\textbf{Multi-turn cascades:}} Gradual cascades arise when individual turn-level signals are not sufficient
for immediate confirmation, but spectral coupling persists across turns. When
\textsc{Watch} first activates at turn $t_w$ and the instant rule is not
satisfied, CASPIAN initiates an adaptive persistence interval. The persistence window, $W_{t_w}$ is governed by the mixing time, bounded by the inverse spectral gap~\cite{levin2017markov}.
We define 
\[W_{t_w} = \left\lceil \frac{1}{g_{t_w}+\epsilon} \right\rceil\] 
The confirmation interval is $[t_w,t_0]$, where $t_0 = t_w + W_{t_w}-1$.

Hence, a \textit{multi-turn cascade} is declared when \textsc{Watch} holds for at least half of the interval and the interval contains at least one \textsc{PhaseShift} or \textsc{CrossChannel} event:

\begin{empheq}[box=\fbox]{equation}
\mathrm{\textsc{MultiTurnCascade}}(t_0)
=
\Bigl[
\mathrm{\textsc{MajorityWatch}}
\;\wedge\;
\mathrm{Transition}
\Bigr](t_w,W_{t_w})
\end{empheq}

Here, \textsc{MajorityWatch} denotes that \textsc{Watch} holds for at least half
of the turns in $[t_w,t_0]$, while \textsc{Transition} denotes the presence of
at least one \textsc{PhaseShift} or \textsc{CrossChannel} event within the
interval. If \textsc{Watch} drops before the confirmation turn, the candidate
interval is discarded and reinitialized at the next \textsc{Watch} onset.

The final cascade decision is 
\[\mathrm{Cascade}(t_0)
=
\mathrm{\textsc{InstantCascade}}(t_0)
\;\vee\;
\mathrm{\textsc{MultiTurnCascade}}(t_0)\]
This formulation separates transient activity bursts from true cascades by requiring early spectral onset to receive either immediate structural confirmation or sustained temporal support.

\subsection{Cascade Attribution via Spectral Flow Decomposition}

Upon detection at turn $t_0$, CASPIAN performs online attribution directly from the influence matrices cached during detection, requiring no additional 
parameters or recomputation. Attribution operates over the onset snapshot $\tilde{A}_{t_w}$ together with 
the retained interval $\{\tilde{A}_\tau, A_\tau\}_{\tau=t_w}^{t_0}$, where 
$t_w$ denotes the first \textsc{Watch} activation turn. For immediate cascades, 
$t_w=t_0$ and attribution reduces to a single-turn snapshot. We identify three agent roles together with the dominant propagation spines of the cascade.

The \textbf{Origin} corresponds to the agent with the strongest outgoing influence at 
cascade onset:
\begin{equation}
i^{\mathrm{origin}}
=
\arg\max_i
\sum_j \tilde{A}_{t_w}(i,j)
\end{equation}
The \textbf{Amplifier} corresponds to the agent that most strongly reinforces 
propagation relative to what it receives across the confirmation interval $[t_w, t_0]$:
\begin{equation}
i^{\mathrm{amp}}
=
\arg\max_i
\sum_{\tau=t_w}^{t_0}
\frac{
\sum_j \tilde{A}_\tau(i,j)
}{
\sum_k \tilde{A}_\tau(k,i)+\epsilon
}
\end{equation}
The \textbf{Bridge} is the agent that most strongly receives and redistributes influence across the network. To avoid suppression of hub-adjacent edges, it is computed using 
the unnormalized matrices $A_\tau$:
\begin{equation}
i^{\mathrm{bridge}}
=
\arg\max_i
\sum_{\tau=t_w}^{t_0}
\left(
\sum_j A_\tau(i,j)
\right)
\left(
\sum_k A_\tau(k,i)
\right)
\end{equation}
We then extract the \textit{\textbf{top-K cascade propagation spines}} from the elementwise maximum influence matrix over the confirmation interval 

\[\bar{A}^{\max}(i,j) = \max_{\tau \in [t_w,\, t_0]}\, \tilde{A}_\tau(i,j)\]

which captures the strongest influence each edge achieves during the cascade rather than its state at any single turn. Spines are then ranked by weakest-link strength:
\begin{equation}
\Pi^K = \operatorname{TopK}_{\pi\,:\,|\pi|\leq \mathrm{diam}(G_0)}\;
\min_{(i,j)\in\pi}\, \bar{A}^{\max}(i,j)
\end{equation}
For each propagation spine $\pi_k\in\Pi^K$, the dominant interaction channel 
is identified as: $c^\star(\pi_k)
=
\arg\max_{c\in\mathcal{C}}
\sum_{(i,j)\in\pi_k}
\bar{A}^{\max,(c)}(i,j)$,
where $\bar{A}^{\max,(c)}$ denotes the maximum interval influence matrix for 
channel $c$ over $[t_w,t_0]$. 

The final attribution output is given by:

\[\Bigl\{
i^{\mathrm{origin}},
\;
i^{\mathrm{amp}},
\;
i^{\mathrm{bridge}},
\;
\Pi^K,
\;
(c^\star(\pi_k))_{k=1}^{K}
\Bigr\}\]

which provides a trace-conditioned influence decomposition of where the cascade originates, how it amplifies, its propagation pathways, and channels.

%% file: sec/6_experiments.tex
\vspace{-6pt}
\section{Experiments}
\label{sec:experiments}

\vspace{-6pt}
\subsection{Experimental Setup}
\label{sec:exp-setup}

\noindent \textbf{Benchmarks and Attack Types:}
We evaluate CASPIAN on two recent benchmarks for cascade failures in LLM multi-agent systems. \textbf{TAMAS}~\cite{kavathekar2025tamas} contains adversarial scenarios spanning intent manipulation, tool and memory-mediated execution attacks, and malicious inter-agent synchronization. \textbf{ACIArena}~\cite{an2026aciarena} evaluates cascading injection attacks across diverse MAS topologies and coordination patterns. Across both benchmarks, we evaluate \textbf{727 scenarios} yielding \textbf{2,908 framework-specific traces}, including \textbf{169 benign} and \textbf{558 attack} scenarios. We group attacks into three cascade categories: \textit{intent manipulation} (communication-driven semantic drift), \textit{execution attacks} (tool and memory propagation), and \textit{coordination attacks} (synchronized multi-agent failure). See Appendix~\ref{supp-sec:benchmarks} for detailed category mappings, trace counts, and confidence intervals.

\noindent \textbf{MAS Frameworks:}
We evaluate CASPIAN across four representative LLM-based multi-agent frameworks: \textbf{AutoGen}~\cite{wu2024autogen}, \textbf{CrewAI}~\cite{CrewAI2024}, \textbf{MetaGPT}~\cite{hong2023metagpt}, and \textbf{LLM Debate}~\cite{liang2024encouraging}. These frameworks span hub-and-spoke, hierarchical, role-specialized, and decentralized interaction structures. Unless otherwise stated, all experiments use \textbf{GPT-5.4}. Results for alternative LLMs are shown in Table~\ref{tab:llm_ablation}.

\noindent \textbf{Baselines:}
We compare CASPIAN against three categories of defenses: \textit{semantic guardrails}, including PromptGuard~2~\cite{chennabasappa2025llamafirewall}, JailGuard~\cite{zhang2025jailguard}, PromptArmor~\cite{shi2025promptarmor}, and perplexity-based detection~\cite{radford2019language}; \textit{LLM-based judges}, including single-turn, sliding-window, and full-trace variants; and \textit{graph-based detectors}, including G-Safeguard~\cite{wang2025g}, BlindGuard~\cite{miao2025blindguard}, and GUARDIAN~\cite{zhou2026guardian}.

\noindent \textbf{Metrics:}
For cascade detection, we report \textbf{AUROC}, \textbf{TPR@5\%FPR}, and \textbf{EDR@5} (Early Detection Rate within 5 turns of injection), which measures whether a cascade is identified before substantial propagation occurs. For attribution, we report \textbf{Acc@1} and \textbf{MRR} for origin, amplifier, and bridge recovery, \textbf{Spine Jaccard@3} for propagation-path recovery, \textbf{Channel Accuracy} for dominant channel recovery, and \textbf{Attribution Lag} measured in turns. Metric definitions are provided in Appendix~\ref{supp-sec:benchmarks}.

\noindent \textbf{Experimental Protocol:}
For each benchmark scenario, the target MAS framework is executed under benign and attack settings, producing multi-turn traces spanning communication, memory, tool, and execution interactions. CASPIAN integrates through a lightweight adapter layer that converts framework-specific logs into a unified cross-channel trace. All online methods process traces strictly turn-by-turn using only past interaction history, while offline baselines receive the full execution trace. Detection is declared at the first firing turn, and EDR@5 measures whether detection occurs within five turns of attack injection. For attribution, CASPIAN recovers cascade roles and propagation paths directly from cached influence dynamics without replay or recomputation. All experiments are conducted on a single NVIDIA A100 GPU; implementation details are provided in Appendix~\ref{supp-sec:benchmarks}.

\vspace{-6pt}
\subsection{Results}

\subsubsection{Cascade Detection Evaluation}

Table~\ref{tab:detection-quant} reports cascade detection performance across benchmarks, attack categories, and MAS frameworks. CASPIAN achieves consistently strong performance across all settings, with AUROC exceeding 0.9 in most configurations and strong EDR@5, indicating that many cascades are detected within only a few turns of injection. This arises from jointly monitoring communication, memory, tool, and execution influence pathways online, allowing weak signals in one channel to be reinforced through correlated propagation structure in others. 

\begin{table*}[t]
\centering
\scriptsize
\setlength{\tabcolsep}{2.6pt}
\renewcommand{\arraystretch}{1.12}
\caption{\textbf{Cascade detection across benchmarks and MAS frameworks.}
We report AUROC$\uparrow$, TPR@5\%FPR$\uparrow$, and early detection rate within 5 turns (EDR@5$\uparrow$). Results are aggregated over attacks grouped by propagation mechanism. LLM Debate exhibits a principled exception where intent manipulation is most detectable due to its communication-dominant architecture. EDR@5 for execution attacks forms a conservative lower bound, requiring several turns to accumulate sufficient influence signal, particularly in MetaGPT and LLM Debate.}
\label{tab:detection-quant}
\vspace{6pt}
\resizebox{\textwidth}{!}{
\begin{tabular}{llccccccccc}
\toprule
& & \multicolumn{3}{c}{\textbf{Intent Manipulation}}
& \multicolumn{3}{c}{\textbf{Execution Attacks (Tools + Memory)}}
& \multicolumn{3}{c}{\textbf{Coordination Attacks}} \\
\cmidrule(lr){3-5} \cmidrule(lr){6-8} \cmidrule(lr){9-11}
\textbf{Bench.} & \textbf{MAS}
& AUROC & TPR@5\% & EDR@5
& AUROC & TPR@5\% & EDR@5
& AUROC & TPR@5\% & EDR@5 \\
\midrule

\multirow{4}{*}{TAMAS}
& AutoGen
& 0.932 & 0.846 & 0.781
& 0.949 & 0.872 & 0.817
& 0.956 & 0.888 & 0.831 \\

& CrewAI
& 0.914 & 0.807 & 0.747
& 0.934 & 0.842 & 0.782
& 0.951 & 0.865 & 0.795 \\

& MetaGPT
& 0.887 & 0.748 & 0.664
& 0.901 & 0.771 & 0.684
& 0.907 & 0.782 & 0.690 \\

& LLM Debate
& 0.894 & 0.761 & 0.793
& 0.872 & 0.726 & 0.748
& 0.881 & 0.741 & 0.774 \\

\midrule

\multirow{4}{*}{ACIArena}
& AutoGen
& 0.918 & 0.813 & 0.758
& 0.946 & 0.858 & 0.804
& 0.948 & 0.869 & 0.814 \\

& CrewAI
& 0.901 & 0.782 & 0.724
& 0.927 & 0.821 & 0.761
& 0.936 & 0.843 & 0.774 \\

& MetaGPT
& 0.872 & 0.723 & 0.631
& 0.889 & 0.754 & 0.654
& 0.893 & 0.762 & 0.661 \\

& LLM Debate
& 0.879 & 0.742 & 0.768
& 0.858 & 0.718 & 0.724
& 0.864 & 0.731 & 0.751 \\

\bottomrule
\end{tabular}
}
\end{table*}

Performance is strongest in AutoGen, where CASPIAN achieves EDR@5 values up to
0.831, while MetaGPT and LLM Debate remain the most challenging settings,
particularly for execution attacks, where EDR@5 drops to 0.684 and 0.661,
respectively. These differences reflect framework topology: AutoGen and CrewAI
exhibit more centralized interaction flow, allowing coordination attacks to be
detected early because synchronization rapidly produces coupled spectral
structure. In contrast, MetaGPT and LLM Debate distribute influence across
specialized roles, memory reuse, and longer interaction horizons, causing
tool- and memory-mediated execution cascades to accumulate more gradually before
sufficient amplification and spectral coupling emerge. LLM Debate exhibits a
distinct pattern in which intent manipulation remains highly detectable because
debate coordination is communication-dominant. We report 95\% bootstrap
confidence intervals over 1,000 resamples in
Appendix~\ref{app:confidence_intervals}.

\noindent \textbf{Comparison with Baselines:} Table~\ref{tab:baselines} compares CASPIAN against 
existing defenses across AutoGen and LLM Debate. Semantic 
guardrails and perplexity-based methods achieve the lowest 
performance across all metrics: PromptGuard~2 reaches 
0.284 TPR@5\% and 0.198 EDR@5 on AutoGen, while 
perplexity-based detection reaches 0.132 TPR@5\% and 
0.076 EDR@5. These methods operate primarily at the level 
of individual messages or token distributions and 
therefore capture limited information about how influence 
propagates across agents and turns.

LLM judges improve as the evaluation window increases, 
rising from 0.412 TPR@5\% and 0.215 EDR@5 for single-turn 
evaluation to 0.612 and 0.441 for sliding-window 
evaluation ($W=10$). However, fixed context windows remain 
sensitive to the propagation timescale of the attack: 
narrow windows miss slower cascades, while larger windows 
delay responsiveness to fast transitions. Full-trace 
evaluation achieves the strongest AUROC among judges 
(0.811 on AutoGen) but does not support early online 
detection.

Graph-based detectors provide the strongest competing 
baselines, with BlindGuard reaching 0.645 TPR@5\%, 0.521 
EDR@5, and 0.815 AUROC on AutoGen. This indicates that 
structural reasoning over agent interactions captures 
important cascade information beyond localized semantic 
inspection. However, their early-detection performance 
remains consistently below CASPIAN across both 
frameworks. CASPIAN achieves 0.868 TPR@5\%, 0.792 EDR@5, 
and 0.942 AUROC on AutoGen, while maintaining similar 
relative improvements on LLM Debate. Performance is lower 
for all methods on LLM Debate, likely due to its more 
distributed and communication-dominant interaction 
structure, which reduces the clarity of cross-channel 
propagation signals.

\begin{table}[t]
\caption{\textbf{Baseline comparison with CASPIAN.} We report true positive rate at 5\% false positive rate (TPR@5\%$\uparrow$), early detection rate within 5 turns (EDR@5$\uparrow$), and AUROC$\uparrow$. EDR is not applicable for full-trace (offline) methods.}
\vspace{6pt}
\label{tab:baselines}
\centering
\small
\begin{tabular}{@{}l rcc c rcc@{}}
\toprule
& \multicolumn{3}{c}{\textbf{AutoGen}} & \phantom{abc} & \multicolumn{3}{c}{\textbf{LLM Debate}} \\
\cmidrule{2-4} \cmidrule{6-8}
\textbf{Method} & TPR@5\% & EDR@5 & AUROC && TPR@5\% & EDR@5 & AUROC \\
\midrule
\textbf{Guardrails} \\
Prompt Guard 2~\cite{chennabasappa2025llamafirewall} & 0.284 & 0.198 & 0.618 && 0.233 & 0.154 & 0.579 \\
JailGuard~\cite{zhang2025jailguard}                 & 0.326 & 0.242 & 0.630 && 0.271 & 0.166 & 0.586 \\
PromptArmor~\cite{shi2025promptarmor}               & 0.361 & 0.251 & 0.664 && 0.304 & 0.207 & 0.622 \\
\addlinespace[0.3em]
\textbf{Distributional} \\
Perplexity~\cite{radford2019language}               & 0.132 & 0.076 & 0.578 && 0.086 & 0.045 & 0.533 \\
\addlinespace[0.3em]
\textbf{LLM Judges} \\
Single Turn                                         & 0.412 & 0.215 & 0.694 && 0.383 & 0.190 & 0.661 \\
Sliding Window ($W=5$)                              & 0.528 & 0.336 & 0.742 && 0.496 & 0.314 & 0.717 \\
Sliding Window ($W=10$)                             & 0.612 & 0.441 & 0.763 && 0.575 & 0.395 & 0.736 \\
Full Trace                                          & 0.713 & ---   & 0.811 && 0.686 & ---   & 0.765 \\
\addlinespace[0.3em]
\textbf{Graph-Based} \\
G-Safeguard~\cite{wang2025g}                        & 0.616 & 0.482 & 0.802 && 0.582 & 0.437 & 0.785 \\
BlindGuard~\cite{miao2025blindguard}                & \underline{0.645} & \underline{0.521} & \underline{0.815} && \underline{0.612} & \underline{0.468} & \underline{0.802} \\
GUARDIAN~\cite{zhou2026guardian}                    & 0.632 & 0.498 & 0.818 && 0.598 & 0.452 & 0.796 \\
\midrule
\textbf{CASPIAN (Ours)}                             & \textbf{0.868} & \textbf{0.792} & \textbf{0.942} && \textbf{0.842} & \textbf{0.753} & \textbf{0.915} \\
\bottomrule
\end{tabular}
\end{table}

\subsubsection{Spectral Dynamics of Cascade Formation}

\begin{figure*}[t]
    \centering
\includegraphics[width=\linewidth]{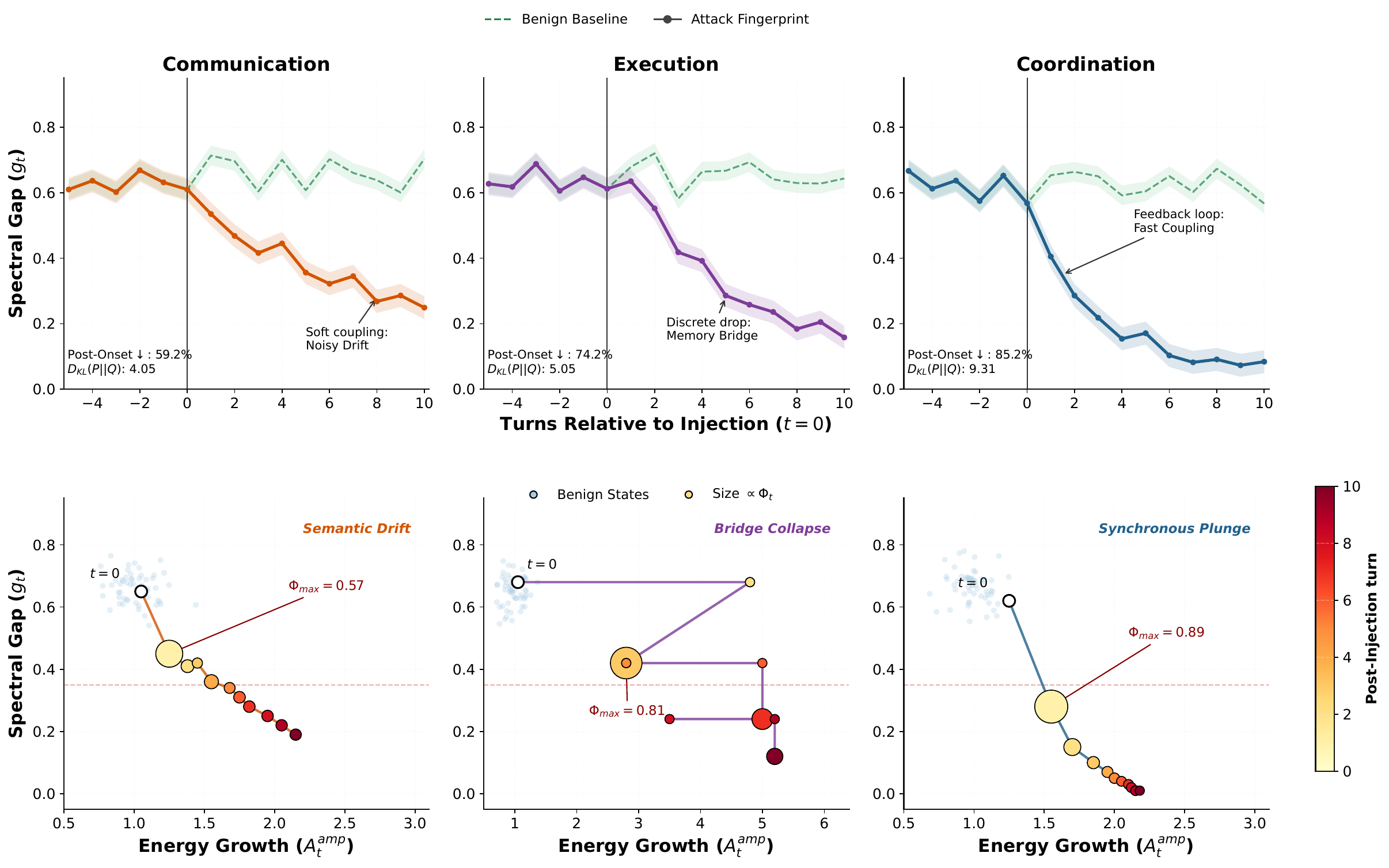}
    \caption{\textbf{Temporal and phase-space dynamics of cascade attacks.}
    \label{fig:spectral-amp-gap-dynamics}
    \textbf{Top:} Spectral gap $g_t$ over turns ($t=0$ denotes injection). Benign trajectories remain in a high-gap regime, while attacks exhibit distinct collapse patterns: gradual drift (communication), stepwise drops (execution), and rapid collapse (coordination).
    \textbf{Bottom:} Phase-space trajectories in $(A_t^{amp}, g_t)$. Marker size $\propto \Phi_t$ (phase shift) and color indicates post-injection progression. The largest $\Phi_t$ occurs near onset ($t=0$), while secondary peaks capture structural transitions (e.g., execution bridges).}
    \label{fig:spectral-gap-amplification}
\end{figure*}

\noindent \textbf{Temporal and Phase-Space Dynamics of Cascade Attacks:} Figure~\ref{fig:spectral-amp-gap-dynamics} presents two 
complementary views of cascade propagation dynamics across 
communication, execution, and coordination attack types: 
the top row tracks the spectral gap $g_t$ over turns 
relative to injection ($t=0$), while the bottom row shows 
corresponding phase-space trajectories in the 
$(A_t^{\mathrm{amp}}, g_t)$ plane, where marker size 
reflects phase-shift magnitude $\Phi_t$ and color 
indicates post-injection turn progression. Benign 
trajectories maintain a consistently high spectral gap 
throughout, confirming that normal collaborative activity 
does not produce sustained spectral compression even when 
interaction intensity fluctuates transiently.

The three attack classes produce qualitatively distinct 
collapse signatures despite sharing similar benign 
pre-injection behavior. Communication attacks exhibit 
gradual semantic drift, with amplification and spectral 
coupling evolving progressively through repeated 
conversational reinforcement, producing a post-onset gap 
reduction of 59.2\% and the lowest divergence from benign 
dynamics ($D_{KL}=4.05$) among the three attack types. 
The corresponding phase-space trajectory follows a smooth 
diffusion-like path, indicating that coupling emerges 
progressively as semantic influence repeatedly circulates 
through the interaction topology.

Execution attacks instead exhibit a two-stage propagation 
pattern in which amplification rises sharply before a 
discrete spectral collapse occurs, reflecting delayed 
synchronization through memory reuse and tool-mediated 
interaction bridges. This produces an intermediate gap 
reduction of 74.2\% ($D_{KL}=5.05$). In phase space, this 
appears as an abrupt directional transition rather than 
gradual drift, indicating that execution cascades remain 
partially localized until delayed synchronization pathways 
activate.

Coordination attacks show the fastest collapse dynamics, 
where amplification growth and synchronization emerge 
nearly simultaneously, driving rapid movement into a 
persistent low-gap regime with the most severe gap 
reduction of 85.2\% and the largest divergence from 
benign behavior ($D_{KL}=9.31$). This behavior is further 
supported by the largest observed phase shift 
$\Phi_{\max}=0.89$ occurring near onset, indicating rapid 
feedback-loop coupling across agents rather than delayed 
propagation through intermediate interaction pathways.

The phase-space trajectories further reveal a structural 
property not directly visible in the temporal plots: the 
largest phase-shift magnitudes $\Phi_t$, indicated by 
the largest markers, occur near attack onset across all 
three attack types, while smaller secondary peaks appear 
later during structural reorganizations such as delayed 
execution-bridge synchronization. This suggests that 
cascade initiation is driven by abrupt reorganization of 
the influence topology near onset, whereas persistent 
propagation is sustained through subsequent reinforcement 
of the coupled spectral structure. The result further 
explains why CASPIAN's phase-shift signal supports early 
online detection: it fires during the initial structural 
transition rather than after the system has already 
entered a fully synchronized cascade regime.

\begin{figure*}[t]
    \centering
\includegraphics[width=1\linewidth]{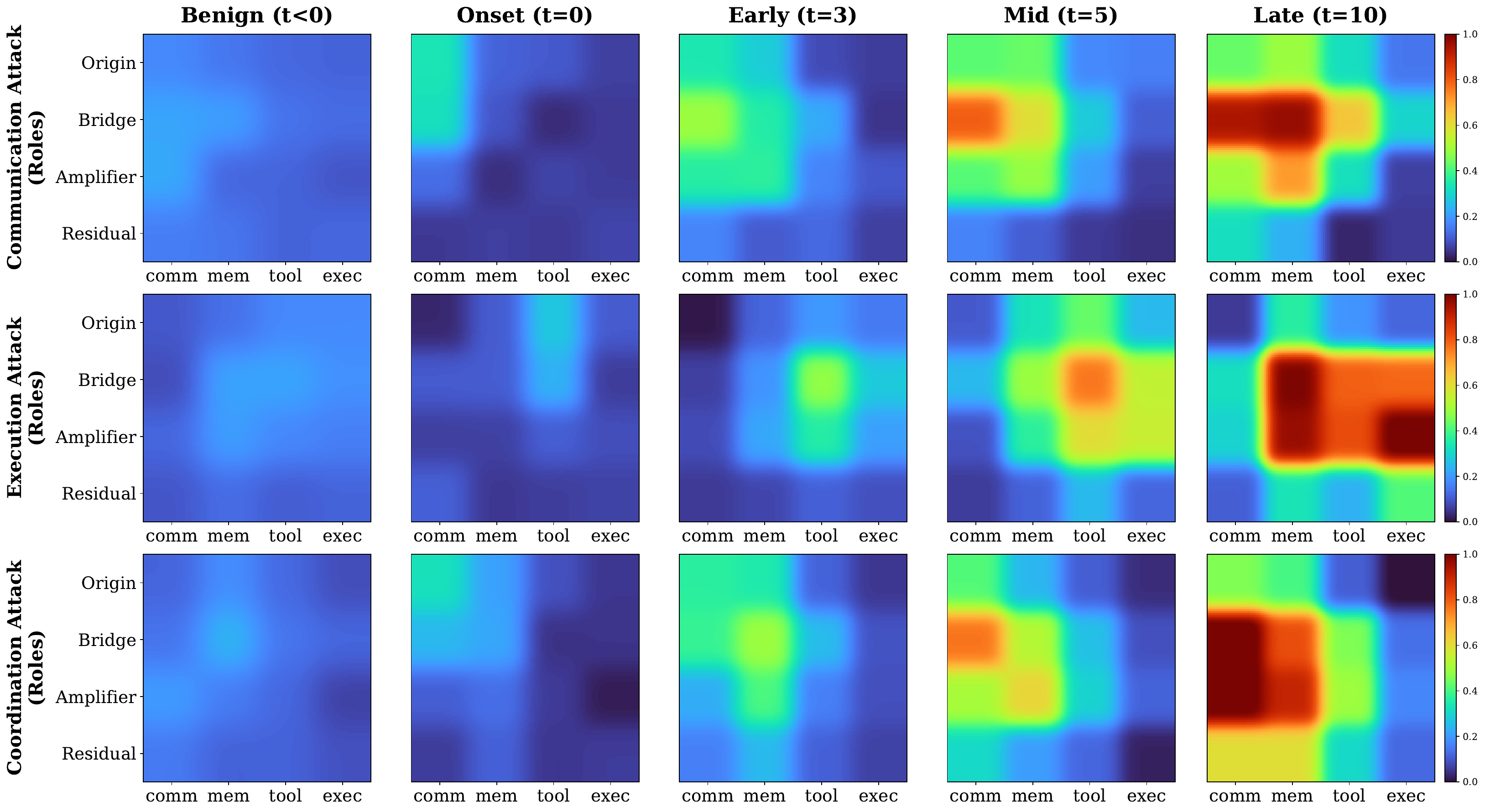}
    \caption{\textbf{Role-channel propagation intensity across 
    cascade attack types.} Each cell shows normalized 
    cross-channel activation intensity (comm, mem, tool, exec) 
    across agent roles (Origin, Bridge, Amplifier, Residual) 
    at five temporal stages: benign ($t<0$), onset ($t=0$), 
    early ($t=3$), mid ($t=5$), and late ($t=10$) 
    post-injection turns. Benign states exhibit uniformly low, 
    diffuse activation across all roles and channels. At 
    onset, origin agents activate first, followed by bridge 
    agents carrying influence toward downstream roles, while 
    amplifier activation broadens and intensifies through mid 
    and late stages. The dominant interaction channel differs 
    systematically by attack type: communication attacks 
    propagate primarily through comm before spreading into 
    mem; execution attacks concentrate sharply in tool and 
    exec channels around bridge roles; coordination attacks 
    activate broadly across all four channels simultaneously, 
    consistent with their rapid cross-channel synchronization.}
    \label{fig:spectral-heatmap}
\end{figure*}

\noindent \textbf{Cross-Channel Propagation Dynamics:} Figure~\ref{fig:spectral-heatmap} visualizes the 
joint evolution of cross-channel activation intensity 
across agent roles and interaction channels at five 
temporal stages spanning benign behavior through late 
cascade propagation. During the benign phase ($t<0$), 
activation remains uniformly low and diffuse across all 
role-channel combinations, with no persistent 
concentration in any agent role or interaction modality, 
confirming that normal collaborative activity does not 
produce the structured role-ordered activation patterns 
characteristic of cascades.

At onset ($t=0$), origin agents exhibit the earliest and 
most concentrated activation, consistent with their role 
as the initial injection point. Bridge agents activate 
subsequently, carrying influence toward downstream regions 
of the topology, while amplifier roles show the broadest 
and most persistent activation through mid and late 
propagation stages. Residual agents activate last and most 
diffusely, reflecting passive receipt of influence rather 
than active propagation. This role-ordered activation 
sequence: origin, bridge, amplifier, residual, is 
consistent across all three attack types despite their 
distinct spectral signatures, suggesting that it reflects 
a structural property of cascade formation rather than 
attack-specific behavior.

The interaction channels through which this sequence 
unfolds differ systematically across attack types. 
Communication attacks concentrate activation in comm 
channels during early propagation before gradually 
spreading into mem interactions at later stages, 
consistent with semantic drift accumulating through 
repeated conversational reinforcement. Execution attacks 
instead exhibit sharp concentration in tool and exec 
channels around bridge roles, reflecting delayed 
synchronization through tool-mediated interaction 
pathways. Coordination attacks activate broadly across all 
four channels within only a few turns after injection, 
producing the strongest cross-channel activation and most 
rapid amplifier emergence, consistent with immediate 
feedback-loop coupling dynamics.

These role-channel activation patterns indicate 
that cascade propagation has both a structural dimension, which agents carry influence, and a modal dimension through which interaction pathways propagation occurs, highlighting the importance of joint cross-channel 
causal modeling for online cascade detection and 
attribution.

\begin{figure*}[t]
\centering
\includegraphics[width=\textwidth]{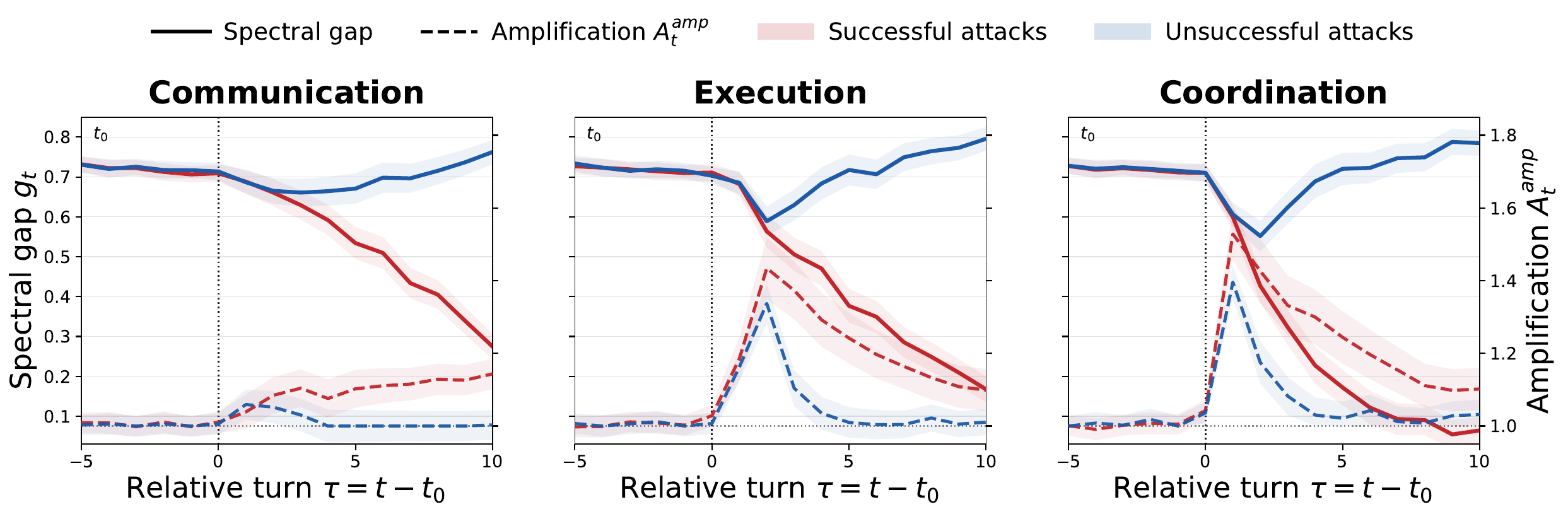}
\caption{\textbf{Spectral dynamics of successful vs.\ unsuccessful cascade attacks by type}, averaged over traces with $\pm1$ std.\ bands. Spectral gap $g_t$ (solid, left axis) and amplification $A_t^{\mathrm{amp}}$ (dashed, right axis) are shown for successful (red) and unsuccessful (blue) attacks over the first 10 post-injection turns. Successful cascades sustain low-gap, high-amplification states, whereas unsuccessful attacks recover within a few turns. Communication attacks produce gradual drift with delayed amplification growth, execution attacks exhibit discrete gap collapse with early amplification spikes, and coordination attacks show the fastest collapse with immediate amplification peaks at $t_0$.}
\label{fig:success-failure-dynamics}
\end{figure*}

\noindent \textbf{Successful vs Unsuccessful Attack Dynamics:} Figure~\ref{fig:success-failure-dynamics} compares successful and unsuccessful cascade attacks through spectral gap and amplification dynamics. Pre-injection trajectories remain nearly indistinguishable, suggesting that cascade signals emerge primarily from the structural response to injection. Successful cascades maintain low-gap, high-amplification states for many turns, indicating persistent synchronization and reinforcement, whereas unsuccessful attacks produce only transient perturbations before returning to stable high-gap structure. This separation is strongest for coordination attacks, while execution attacks exhibit intermediate behavior depending on whether propagation spreads successfully through memory and tool pathways.

\vspace{-6pt}
\subsubsection{Online Cascade Attribution}
\vspace{-6pt}
Table~\ref{tab:attribution_eval} evaluates CASPIAN's ability 
to localize cascade roles and recover propagation 
structure at detection time across two benchmarks and 
three MAS frameworks. Ground-truth attribution labels are derived from benchmark attack execution traces, including injection sources, intermediate relay agents, and observed propagation paths 
across interaction channels (See Appendix~\ref{supp-sec:benchmarks}). Origin, amplifier, and bridge 
Acc@1 measure the fraction of cascades for which the 
highest-ranked predicted role matches the ground-truth 
agent, while MRR evaluates how highly the correct agent is 
ranked on average across all predictions. Origin 
localization is strongest overall, reaching 0.860 Acc@1 
and 0.911 MRR on TAMAS AutoGen, indicating that agents 
producing the strongest outgoing causal influence are 
identifiable from the evolving influence topology during 
online execution. Amplifier recovery is similarly stable, 
with Acc@1 values above 0.75 across most settings, 
suggesting that the outflow-to-inflow ratio computed over 
the confirmation window provides a consistent amplifier 
signal under online constraints.

Bridge localization is comparatively more difficult across 
all settings, with Acc@1 values ranging from 0.680 to 
0.734 on TAMAS and 0.650 to 0.720 on ACIArena. Bridge 
agents both receive and redistribute large volumes of 
influence, making them harder to separate from 
high-connectivity hub agents that accumulate interaction 
volume due to structural position alone. CrewAI exhibits 
slightly stronger bridge recovery than AutoGen despite 
weaker origin and amplifier localization, likely because 
its hub-and-spoke coordination structure concentrates 
relay behavior into a smaller number of identifiable 
agents. In contrast, LLM Debate produces the weakest 
bridge localization across both benchmarks, consistent 
with its more distributed, communication-heavy interaction 
structure where relay pathways are less centralized.

\begin{table*}[t]
\centering
\small
\setlength{\tabcolsep}{5pt}
\renewcommand{\arraystretch}{1.12}
\caption{\textbf{Cascade attribution across benchmarks and MAS frameworks.}
We evaluate role localization (Origin, Amplifier, Bridge) using Acc@1 and MRR, joint role identification using Joint Acc@1, and propagation recovery using Spine Jaccard@3 and Channel Accuracy.
Results are reported on successful cascades at detection time.
Bridge identification is slightly stronger in CrewAI due to its hub-and-spoke topology, where relay structure is more centralized and thus more easily captured by the accumulated in-out flow criterion.
Attribution Lag (in turns) measures the delay between cascade onset and confirmation.}
\label{tab:attribution_eval}
\vspace{6pt}
\begin{tabular*}{\textwidth}{@{\extracolsep{\fill}}lcccccc}
\toprule
\multirow{2}{*}{\textbf{Metric}} 
& \multicolumn{3}{c}{\textbf{TAMAS}} 
& \multicolumn{3}{c}{\textbf{ACIArena}} \\
\cmidrule(lr){2-4} \cmidrule(lr){5-7}
& \textbf{AutoGen} & \textbf{CrewAI} & \textbf{LLM Debate}
& \textbf{AutoGen} & \textbf{CrewAI} & \textbf{LLM Debate} \\
\midrule

\multicolumn{7}{l}{\textbf{Role Localization}} \\
Origin Acc@1        & 0.860 & 0.794 & 0.743 & 0.832 & 0.789 & 0.728 \\
Origin MRR          & 0.911 & 0.876 & 0.826 & 0.872 & 0.845 & 0.802  \\
Amplifier Acc@1     & 0.807 & 0.756 & 0.722 & 0.778 & 0.723 & 0.689 \\
Amplifier MRR       & 0.843 & 0.819 & 0.780 & 0.827 & 0.779 & 0.752 \\
Bridge Acc@1        & 0.734 & 0.725 & 0.680 & 0.710 & 0.720 & 0.650 \\
Bridge MRR          & 0.816 & 0.822 & 0.755 & 0.783 & 0.794 & 0.728 \\
Joint Acc@1         & 0.509 & 0.435 & 0.364 & 0.459 & 0.410 & 0.326 \\

\midrule
\multicolumn{7}{l}{\textbf{Propagation Recovery}} \\
Spine Jac.@3        & 0.733 & 0.694 & 0.651 & 0.708 & 0.664 & 0.626 \\
Channel Acc         & 0.892 & 0.866 & 0.837 & 0.860 & 0.836 & 0.802 \\
Attribution Lag     & 1.451 & 1.726 & 2.355 & 1.810 & 2.101 & 2.706 \\

\bottomrule
\end{tabular*}
\end{table*}

Joint Acc@1 measures whether all three roles: origin, 
amplifier, and bridge, are simultaneously identified 
correctly for a cascade. Scores range from 0.509 on TAMAS 
AutoGen to 0.364 on TAMAS LLM Debate, reflecting the 
compounding of individual role estimation errors during 
joint evaluation. Since the metric requires all three role 
predictions to be correct simultaneously, even moderate 
errors in individual role localization produce substantial 
reductions in joint accuracy, particularly in distributed 
frameworks where bridge structure is less centralized.

Propagation recovery metrics further indicate that the 
incrementally maintained influence topology preserves 
meaningful structural information during online 
propagation. Spine Jaccard@3 measures overlap between the 
top-3 recovered propagation paths and the ground-truth 
cascade routes, remaining above 0.62 across all settings. 
Channel Accuracy measures whether the dominant interaction 
modality associated with each recovered spine is correctly 
identified, exceeding 0.80 throughout and reaching 0.892 
on TAMAS AutoGen. Attribution Lag measures the delay 
between cascade onset and successful attribution, 
increasing systematically with framework complexity,  
from 1.451 turns on TAMAS AutoGen to 2.706 on ACIArena 
LLM Debate, reflecting the longer confirmation windows 
required in frameworks where propagation accumulates more 
gradually before stable spectral synchronization emerges.

\vspace{-6pt}
\subsection{Ablation}
\vspace{-6pt}
Table~\ref{tab:ablation} shows ablation of CASPIAN components. Within the influence estimation block, cosine similarity performs worst across all metrics, while mutual information substantially improves performance, confirming the importance of directed, history-aware influence modeling. Full CTE achieves the highest AUROC (0.908) but lower TPR@5\% and EDR@5 than Full CASPIAN, indicating noisy and transient correlations that degrade early detection. LI-CTE suppresses this noise through late-interaction aggregation while preserving strong separability. Channel ablations further show that cascade evidence is distributed across communication, memory, tool, and execution interactions. Comm-only propagation substantially reduces performance, while progressively adding channels consistently improves all metrics. Similarly, using the raw propagation matrix $A_t$ lowers AUROC and TPR@5\%, suggesting that topology normalization stabilizes influence estimation across heterogeneous MAS structures. Within the detection stack, each additional spectral component improves AUROC and EDR@5. Using only $\lambda_1$ provides a coarse amplification signal, while adding gap contraction $\Delta g_t$, phase shift $\Phi_t$, and cross-channel entropy $H_t^{\mathrm{norm}}$ improves sensitivity to synchronization, structural transition, and distributed propagation. Removing WeakLink slightly improves EDR@5 while reducing AUROC and TPR@5\%, due to earlier but noisier detections. Fixed temporal windows fail to adapt to varying cascade timescales. See Appendix~\ref{supp-sec:additional-ablations} for additional ablations.

\begin{table}[t]
\centering
\caption{\textbf{Component Ablation of CASPIAN.} Results are scenario-weighted across benchmarks, MAS frameworks, and attack mechanisms.}
\vspace{6pt}
\label{tab:ablation}
\small
\begin{tabular}{llccc}
\toprule
\textbf{Component} & \textbf{Variant} & \textbf{AUROC} & \textbf{TPR@5\%} & \textbf{EDR@5} \\
\midrule
Influence & Cosine similarity & 0.748 & 0.534 & 0.575 \\
 & MI (no conditioning) & 0.797 & 0.603 & 0.630 \\
 & Full CTE & \textbf{0.908} & 0.752 & 0.696 \\
\addlinespace[0.5em]
Channel & Comm-only & 0.806 & 0.612 & 0.606 \\
 & Comm + Mem + Tool & 0.883 & 0.753 & 0.712 \\
\addlinespace[0.5em]
Normalization & Raw $A_t$ & 0.871 & 0.670 & 0.705 \\
\addlinespace[0.5em]
Detection & $\lambda_1$ only & 0.827 & 0.610 & 0.636 \\
 & + gap contraction $\Delta g_t$ & 0.864 & 0.698 & 0.671 \\
 & + phase shift $\Phi_t$ & 0.882 & 0.734 & 0.711 \\
 & + cross-channel entropy $H_t^{\text{norm}}$ & 0.897 & 0.764 & 0.729 \\
\addlinespace[0.5em]
Path/Temp. & No WeakLink & 0.891 & 0.715 & \textbf{0.746} \\
 & Fixed $W = 5$ & 0.889 & 0.767 & 0.707 \\
\midrule
\textbf{Full CASPIAN} & & \textbf{0.906} & \textbf{0.790} & \textbf{0.743} \\
\bottomrule
\end{tabular}
\end{table}

%% file: sec/7_limitations.tex
\vspace{-6pt}
\section{Discussions and Broader Impact}
\label{supp-sec:broader-impact}
\vspace{-6pt}

CASPIAN shows that cascade attacks in LLM-based multi-agent systems are shaped not only by localized message anomalies, but by the evolving structure of causal influence across agents, channels, and turns. Our results suggest that reliable detection requires joint monitoring of communication, memory, tool, and execution interactions: ablations show that execution-channel evidence cannot be fully recovered from the remaining channels, so message-only monitoring can miss important propagation signals in systems with memory or tool access. Our spectral and role-channel analyses further indicate that, despite differences across attack classes, cascades often follow recurring organizational patterns involving origin, bridge, amplifier, and residual agents. The fact that successful and unsuccessful attacks are largely indistinguishable before injection suggests that vulnerability emerges primarily during online interaction, rather than from static properties of the resting system.

As LLM-MAS are increasingly deployed with persistent memory, tools, and autonomous coordination, failures that propagate across agents may become difficult to localize through message inspection alone. CASPIAN's low relative runtime overhead suggests that cross-channel causal monitoring can be integrated into online MAS execution with limited impact on throughput. Beyond security, tracking influence propagation may support interpretability, auditing, and analysis of agent coordination, motivating further work on structural transparency in complex LLM-agent ecosystems.

\vspace{-6pt}
\section{Limitations}
\label{sec:limitations}
\vspace{-6pt}

CASPIAN assumes access to sufficient interaction traces across communication, memory, tool, and execution channels to construct the evolving causal influence tensor online. In deployments where some modalities are partially observable or unavailable, performance may degrade toward the corresponding channel-ablation setting. However, because LI-CTE models cross-channel propagation jointly, weak evidence in one modality can still be reinforced by correlated structure in other channels. Attribution is also more challenging in highly distributed MAS frameworks such as LLM Debate, where propagation paths are diffuse and bridge roles are less structurally concentrated. Finally, our evaluation focuses on existing MAS frameworks and attack families. Extending CASPIAN to emerging interaction modalities, partially observable deployments, and adaptive adversaries optimized to evade propagation-based monitoring remains an important direction for future work.

\section{Conclusion}
\vspace{-6pt}
We introduced CASPIAN, a unified framework for online detection and attribution of cascade attacks in LLM-based multi-agent systems. CASPIAN combines LI-CTE-based influence estimation with spectral monitoring of amplification, synchronization, persistence, and cross-channel spread, enabling detection of abrupt and gradual cascades without offline training or attack-specific templates. Across benchmarks and MAS frameworks, CASPIAN outperforms semantic guardrails, LLM judges, and graph-based detectors, while also recovering cascade roles and propagation paths during execution. These results highlight structural causal monitoring as a scalable foundation for security, attribution, and auditing in complex LLM-agent ecosystems.

%% file: sec/X_supp.tex
\setcounter{page}{1}
\appendix
\section*{Table of Contents}
\addcontentsline{toc}{section}{Supplementary Material Table of Contents}
\startcontents[appendix]
\printcontents[appendix]{l}{1}{\setcounter{tocdepth}{2}}

\section{Additional Ablations}
\label{supp-sec:additional-ablations}

\subsection{Cross-Benchmark and MAS Ablation Sensitivity}
\label{supp-sec:benchmark-mas-ablation-heatmap-analysis}

\begin{figure*}[t]
    \centering
\includegraphics[width=1\linewidth]{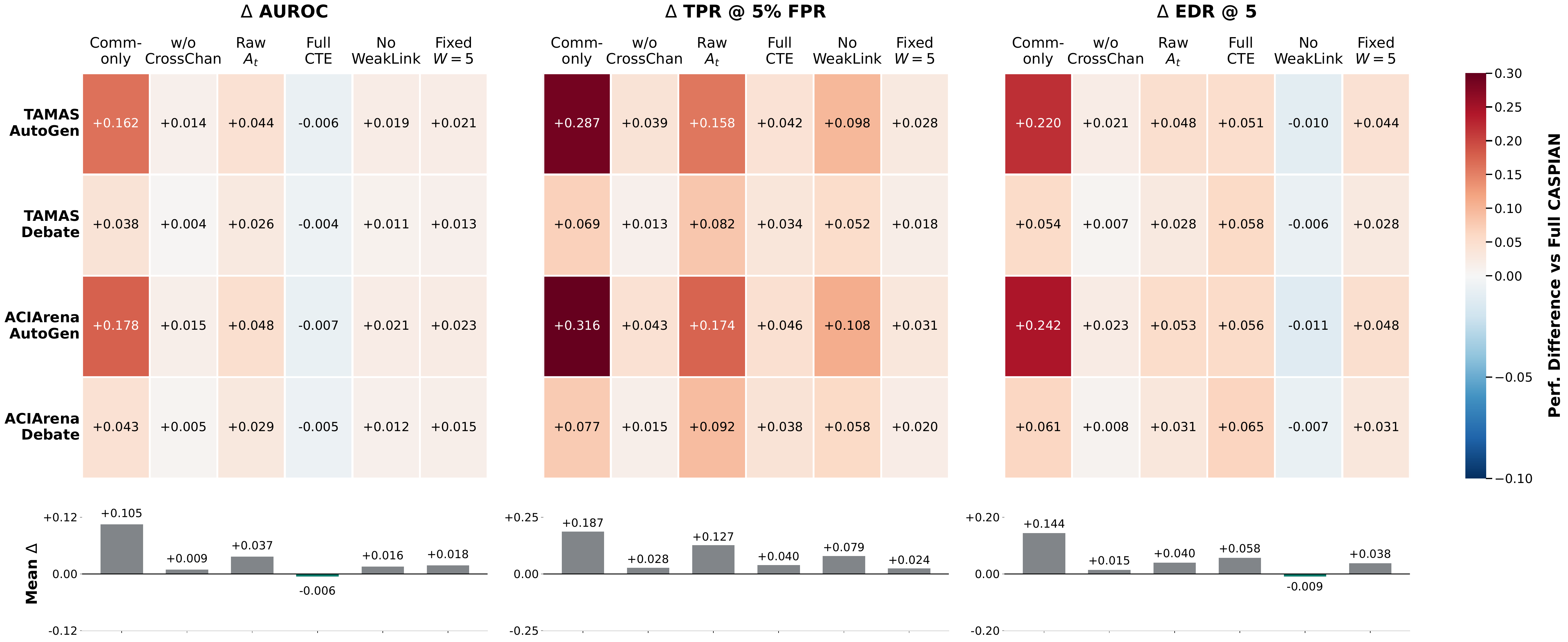}
    \caption{
    \textbf{Ablation sensitivity across benchmarks and MAS frameworks.}
    Cells show signed performance difference ($\Delta$) relative to full CASPIAN (red = degradation, blue = improvement) for AUROC, TPR@5\%FPR, and EDR@5. Restricting to a single channel (\textit{Comm-only}) or removing cross-channel coupling leads to the largest drops, while structural refinements (raw $A_t$, weak-link removal, fixed window) have moderate impact. Bottom bars report mean $\Delta$ across benchmarks.
    }
    \label{fig:ablation-mas-benchmarks}
\end{figure*}

Figure~\ref{fig:ablation-mas-benchmarks} visualizes the 
signed performance difference of each ablation relative to 
Full CASPIAN across benchmarks and MAS frameworks. The 
largest degradations consistently occur when restricting 
the system to a single interaction channel, particularly 
for TPR@5\% and EDR@5, confirming that early cascade 
signals are distributed across communication, memory, 
tool, and execution pathways rather than localized within 
a single modality. Influence estimation ablations also 
produce substantial degradation across nearly all settings, 
especially in AutoGen where propagation structure is more 
explicitly observable. In contrast, structural refinements 
such as topology normalization, \textsc{WeakLink} filtering, and 
adaptive windowing produce more moderate but consistent 
effects, indicating that these components primarily refine 
detection stability and precision once cross-channel 
causal propagation has already been captured.

\paragraph{LLM backbone sensitivity.}
Table~\ref{tab:llm_ablation} evaluates CASPIAN 
across multiple LLM backbones. Detection performance 
remains consistently strong across all models, indicating 
that CASPIAN primarily depends on structural propagation 
dynamics rather than backbone-specific semantic behavior. 
Smaller models such as GPT-4o-mini and Gemini~3.1 show 
moderate degradation in both AUROC and EDR@5, likely due 
to noisier or less stable interaction patterns that reduce 
the clarity of the induced influence topology. In 
contrast, reasoning-oriented or instruction-tuned models 
such as Llama-3.1-8B-Instruct, DeepSeek-R1-Distill, and 
Qwen3-8B maintain performance close to or slightly above 
the GPT-5.4 baseline, suggesting that stronger reasoning 
consistency and more coherent multi-turn interaction 
structure produce cleaner propagation signatures for 
spectral detection.

\subsection{Parameter Sensitivity}

\paragraph{Threshold and persistence sensitivity.}
Table~\ref{tab:threshold_sensitivity} evaluates sensitivity to the
cross-channel entropy threshold $H_t^{\mathrm{norm}}$ and the persistence
window rule $W_t$. We do not tune a \textsc{WeakLink} percentile: CASPIAN's
\textsc{WeakLink} check is parameter-free and compares the widest-path bottleneck
$B_t$ against the graph's energy-weighted influence scale
$\bar w_t^{\,\mathrm{eng}}$. CASPIAN remains stable across moderate
cross-channel entropy thresholds, while the adaptive spectral-gap window
$W_t=\lceil 1/g_t\rceil$ outperforms fixed windows by adapting confirmation
time to the current coupling strength.

\begin{table}[h]
\centering
\caption{\textbf{Threshold and persistence sensitivity.} CASPIAN remains
stable across cross-channel entropy thresholds and benefits from the adaptive
spectral-gap persistence window. \textsc{WeakLink} is not percentile-tuned; it is
defined by the parameter-free energy-weighted bottleneck criterion.}
\vspace{6pt}
\label{tab:threshold_sensitivity}
\small
\setlength{\tabcolsep}{7pt}
\begin{tabular}{llccc}
\toprule
\textbf{Parameter} & \textbf{Setting} & \textbf{AUROC}$\uparrow$ & \textbf{TPR@5\%}$\uparrow$ & \textbf{EDR@5}$\uparrow$ \\
\midrule
\multirow{5}{*}{$H_t^{\mathrm{norm}}$}
& 0.3 & 0.895 & 0.761 & 0.752 \\
& 0.4 & 0.903 & 0.783 & 0.748 \\
& \textbf{0.5} & \textbf{0.906} & \textbf{0.790} & 0.743 \\
& 0.6 & 0.901 & 0.778 & 0.732 \\
& 0.7 & 0.892 & 0.751 & 0.710 \\
\midrule
\multirow{4}{*}{$W_t$}
& $\mathbf{\lceil 1/g_t\rceil}$ & \textbf{0.906} & \textbf{0.790} & \textbf{0.743} \\
& 3 & 0.872 & 0.741 & 0.721 \\
& 5 & 0.889 & 0.767 & 0.707 \\
& 10 & 0.895 & 0.773 & 0.654 \\
\bottomrule
\end{tabular}
\end{table}

\begin{table}[h]
\centering
\small
\setlength{\tabcolsep}{4pt}
\renewcommand{\arraystretch}{1.12}
\caption{\textbf{LLM Ablation of CASPIAN.}
Performance across different LLM backbones. We report AUROC$\uparrow$ and early detection rate within 5 turns (EDR@5$\uparrow$). 
Arrows indicate change relative to GPT-5.4.}
\vspace{6pt}
\label{tab:llm_ablation}
\newcommand{\drop}[1]{\textcolor{pastelorange}{\scriptsize{$\downarrow$#1}}}
\newcommand{\gain}[1]{\textcolor{pastelblue}{\scriptsize{$\uparrow$#1}}}

\begin{tabular*}{\columnwidth}{@{\extracolsep{\fill}}lcccc}
\toprule
\textbf{LLM}
& \multicolumn{2}{c}{\textbf{AutoGen}}
& \multicolumn{2}{c}{\textbf{LLM Debate}} \\
\cmidrule(lr){2-3} \cmidrule(lr){4-5}
& AUROC & EDR@5 & AUROC & EDR@5 \\
\midrule

GPT-5.4 (baseline)
& \textbf{0.942} & \textbf{0.792}
& \textbf{0.875} & \textbf{0.753} \\

GPT-4o-mini 
& 0.902 \drop{0.040}
& 0.733 \drop{0.059}
& 0.834 \drop{0.041}
& 0.701 \drop{0.052} \\

Gemini 3.1 Pro Preview 
& 0.861 \drop{0.081}
& 0.683 \drop{0.109}
& 0.788 \drop{0.087}
& 0.671 \drop{0.082} \\

Llama-3.1-8B-Instruct 
& 0.911 \drop{0.031}
& 0.814 \gain{0.022}
& 0.829 \drop{0.046}
& 0.795 \gain{0.042} \\

DeepSeek-R1-Distill-Llama-8B 
& 0.949 \gain{0.007}
& 0.813 \gain{0.021}
& 0.879 \gain{0.004}
& 0.769 \gain{0.016} \\

Qwen3-8B 
& 0.936 \drop{0.006}
& 0.805 \gain{0.013}
& 0.865 \drop{0.010}
& 0.773 \gain{0.020} \\

\bottomrule
\end{tabular*}
\end{table}

\section{Latency with CASPIAN}
\label{supp-sec:latency}

\paragraph{Runtime overhead.}
Table~\ref{tab:runtime} reports average per-turn runtime 
with and without CASPIAN during online execution on GPT-5.4 
using A100 GPUs. Across both TAMAS and ACIArena, CASPIAN 
introduces only a small additional latency per turn, with 
relative overhead remaining below 1\% in all settings. 
Per-turn runtime is higher for successful attacks due to 
longer propagation chains and denser multi-agent 
interaction patterns, while ACIArena exhibits larger base 
latency overall because of its longer-horizon interaction 
structure. The additional overhead from CASPIAN primarily 
comes from incremental LI-CTE updates, spectral 
decomposition, and cached attribution computations, all of 
which operate over compact influence matrices without 
requiring replay or additional LLM inference.

Per turn, LI-CTE updates scale as $O(|E_0||\mathcal{C}|d)$ for state dimension
$d$, singular-value computation over the compact $N \times N$ influence matrix
costs $O(N^3)$ in the worst case and is negligible for the small-to-moderate
agent teams studied here, and widest-path computation costs
$O(|E_0|\log |V|)$. No additional LLM calls are introduced.

\begin{table}[t]
\centering
\caption{\textbf{Average per-turn runtime overhead of CASPIAN.}
Reported over benign traces and attack traces on TAMAS and
ACIArena using GPT-5.4 on A100 GPUs.}
\vspace{6pt}
\label{tab:runtime}
\small
\begin{tabular}{llcccc}
\toprule
Benchmark & Setting &
Base (s) &
+ CASPIAN (s) &
$\Delta$ (ms) &
Overhead \\
\midrule
TAMAS
& Benign                & 12.8 & 12.9 & 84  & 0.66\% \\
& Successful attacks    & 18.6 & 18.8 & 173 & 0.93\% \\
& Unsuccessful attacks  & 15.1 & 15.2 & 109 & 0.72\% \\
\midrule
ACIArena
& Benign                & 21.4 & 21.6 & 201 & 0.94\% \\
& Successful attacks    & 29.8 & 30.1 & 287 & 0.96\% \\
& Unsuccessful attacks  & 24.2 & 24.4 & 216 & 0.89\% \\
\bottomrule
\end{tabular}
\end{table}

\section{Late-Interaction Conditional Transfer Entropy Computation}
\label{supp-sec:li-cte}

This section describes how CASPIAN converts raw LLM-MAS execution traces into
directed cross-channel influence estimates. The goal of late interaction conditional transfer entropy (LI-CTE), is to estimate,
for each source agent $a_i$, target agent $a_j$, channel $c$, and turn $t$,
whether activity from $a_i$ provides additional information about the current
behavior of $a_j$ beyond what is already explained by $a_j$'s own previous
history. We use ``causal influence'' in the operational sense of directed,
history-conditioned predictive dependence over temporally ordered MAS traces.
CASPIAN does not assume access to counterfactual interventions during execution;
rather, it estimates whether source-agent activity precedes and explains
downstream target behavior after conditioning on the target's prior channel
history and the structural possibility graph.

\noindent\textbf{(1) Normalizing LLM-MAS traces into channel events.}
Native MAS frameworks log interactions in different formats, including chat
messages, role responses, tool calls, memory operations, code execution outputs,
and runtime metadata. CASPIAN first converts these heterogeneous logs into a
unified set of channel events. At turn $t$, the adapter extracts

\begin{equation}
\mathcal{E}_t
=
\{
e_t^{(\mathrm{comm})},
e_t^{(\mathrm{mem})},
e_t^{(\mathrm{tool})},
e_t^{(\mathrm{exec})}
\},
\label{eq:event_trace}
\end{equation}

where each event is represented as
\[
e = (\mathrm{src}, \mathrm{tgt}, c, \mathrm{payload}, \mathbf{x}_e).
\]
Here, $\mathrm{src}$ is the source agent, $\mathrm{tgt}$ is the affected target
agent, $c$ is the interaction channel, $\mathrm{payload}$ is the raw event
content, and $\mathbf{x}_e$ is a compact channel-specific feature vector.
Broadcast or group-chat messages are expanded into multiple directed events
from the sender to every downstream agent that can observe the message. Events
whose endpoints are not feasible under the structural possibility graph $G_0$
are masked out.

\begin{table}[h]
\centering
\caption{\textbf{Operational construction of LI-CTE channel inputs.}
Each framework-specific event is mapped into a directed source--target--channel
signal before influence scoring.}
\vspace{6pt}
\label{tab:licte_operational_channels}
\small
\resizebox{\linewidth}{!}{%
\begin{tabular}{lll}
\toprule
\textbf{Channel} & \textbf{Observed MAS event} & \textbf{LI-CTE signal} \\
\midrule
\texttt{comm}
& Agent message, debate response, role output
& Text embedding, sender/receiver metadata, turn position \\
\texttt{mem}
& Memory read/write, persistent artifact reuse
& Memory-content embedding, operation type, artifact reference \\
\texttt{tool}
& Tool/API call, arguments, returned output
& Tool name, argument features, output representation, call/result type \\
\texttt{exec}
& Runtime behavior during agent execution
& Token usage, latency, completion status, error indicators \\
\bottomrule
\end{tabular}%
}
\end{table}

The meaning of influence differs slightly across channels. In the
communication channel, influence corresponds to whether a source message helps
explain the downstream response of another agent. In the memory channel, it
captures whether a memory artifact written or exposed by one agent later helps
explain the behavior of an agent that reads or reuses that artifact. In the tool
channel, it captures whether a tool invocation or returned output affects
downstream agents. In the execution channel, the signal is not semantic content
but behavioral metadata such as latency, token usage, completion status, or
execution errors. This allows CASPIAN to capture propagation that appears
through runtime behavior even when visible text appears locally benign.

\noindent\textbf{(2) Grouping events by directed edge and channel.}
CASPIAN estimates influence at the level of directed edge-channel triplets
$(i,j,c)$. For each turn $t$, the adapter groups normalized events as

\begin{equation}
\mathcal{B}_{ij}^{(c)}(t)
=
\{e\in \mathcal{E}_t:
\mathrm{src}(e)=i,\;
\mathrm{tgt}(e)=j,\;
c(e)=c\}.
\label{eq:event_bucket}
\end{equation}

If no event occurs for a triplet $(i,j,c)$ at turn $t$, CASPIAN assigns no new
evidence to that edge-channel pair for the turn. If multiple events occur on
the same triplet, their vectors are averaged into a single source-side and
target-side signal:

\begin{equation}
\bar{\mathbf{u}}_{ij}^{(c)}(t)
=
\frac{1}{|\mathcal{B}_{ij}^{(c)}(t)|}
\sum_{e\in\mathcal{B}_{ij}^{(c)}(t)}
\mathbf{u}_e,
\qquad
\bar{\mathbf{v}}_{ij}^{(c)}(t)
=
\frac{1}{|\mathcal{B}_{ij}^{(c)}(t)|}
\sum_{e\in\mathcal{B}_{ij}^{(c)}(t)}
\mathbf{v}_e .
\label{eq:event_average}
\end{equation}

Here, $\mathbf{u}_e$ denotes the source-side representation of the event and
$\mathbf{v}_e$ denotes the target-side representation. For text channels, these
are embedding-based vectors augmented with lightweight metadata. For execution
channels, they are numeric runtime-feature vectors. Averaging multiple events
within a turn reduces local noise while preserving directed source-to-target
channel structure.

\noindent\textbf{(3) Maintaining channel-specific target histories.}
For every target agent $a_j$ and channel $c$, CASPIAN maintains a streaming
history vector $h_j^{(c)}(t)$ summarizing the target's previous behavior in that
channel. Let $\bar{\mathbf{v}}_{j}^{(c)}(t)$ denote the aggregate target-side
state for agent $a_j$ on channel $c$ at turn $t$, obtained by averaging all
target-side vectors associated with events affecting $a_j$ through channel $c$.
The target history is updated using the same channel-agnostic online smoothing
rule for all channels:

\begin{equation}
h_j^{(c)}(t)
=
\mathrm{EMA}
\!\left(
h_j^{(c)}(t-1),
\bar{\mathbf{v}}_{j}^{(c)}(t)
\right).
\label{eq:history_update}
\end{equation}

This history vector is used only as the conditioning context for the target
agent. Importantly, LI-CTE scores the current turn using $h_j^{(c)}(t-1)$, the
target history before observing the current event. The history is updated only
after the influence score is computed, preventing the current target
observation from leaking into the conditioning context. Channel-specific
behavior is captured by the event representations themselves: communication,
memory, tool, and execution events produce different feature signals, rather
than by manually assigning different decay rates to different channels.

\noindent\textbf{(4) Late-interaction conditional influence scoring.}
For a source agent $a_i$, target agent $a_j$, and channel $c$, CASPIAN estimates
directed influence as

\begin{equation}
A_t^{(c)}(i,j)
\propto
I\!\left(
u_i^{(c)}(t);
v_j^{(c)}(t)
\mid
h_j^{(c)}(t-1)
\right),
\label{eq:licte}
\end{equation}

where $u_i^{(c)}(t)$ is the current source-side channel signal,
$v_j^{(c)}(t)$ is the current target-side signal, and
$h_j^{(c)}(t-1)$ is the target's previous channel history.

Operationally, this asks: after accounting for what the target agent was
already likely to do based on its own past behavior, does the source agent's
channel activity explain additional variation in the target's current behavior?
If yes, the edge $(i\rightarrow j)$ receives a high influence score for channel
$c$.

The late-interaction design is important for efficiency. Instead of jointly
encoding full source, target, and history contexts through an expensive
high-dimensional density estimator, CASPIAN constructs compact source, target,
and history representations independently. These representations are combined
only at the final conditional-dependence scoring stage. This preserves directed,
history-aware influence information while avoiding replay, additional LLM
calls, or expensive full-state CTE estimation.

\noindent\textbf{(5) Residual dependence view.}
LI-CTE can be understood as a residual-dependence computation. CASPIAN first
estimates the component of the current target signal that is already explained
by the target's own channel history:

\begin{equation}
\widehat{v}_j^{(c)}(t)
=
P\!\left(
v_j^{(c)}(t)
\mid
h_j^{(c)}(t-1)
\right).
\label{eq:target_prediction}
\end{equation}

It then computes the unexplained target residual:

\begin{equation}
r_j^{(c)}(t)
=
v_j^{(c)}(t)
-
\widehat{v}_j^{(c)}(t),
\label{eq:residual}
\end{equation}

and scores how strongly the source signal depends on this residual:

\begin{equation}
A_t^{(c)}(i,j)
=
\mathrm{Dep}
\!\left(
u_i^{(c)}(t),
r_j^{(c)}(t)
\right).
\label{eq:dep}
\end{equation}

Thus, high LI-CTE does not simply mean that two agents are semantically similar
or active at the same time. It means the source signal explains the target's
current behavior beyond the target's own prior history. This conditioning step
reduces spurious influence caused by shared prompts, common task context, or
background coordination.

\noindent\textbf{(6) Streaming implementation.}
At each turn, CASPIAN updates every feasible edge-channel triplet $(i,j,c)$
through the following procedure:

\begin{enumerate}
    \item \textbf{Collect events:} gather all normalized events from source
    $a_i$ to target $a_j$ through channel $c$.

    \item \textbf{Encode payloads:} convert each payload into a compact
    channel vector. Textual channels use embedding-based representations;
    execution channels use lightweight numeric runtime features.

    \item \textbf{Aggregate within turn:} average multiple events on the same
    $(i,j,c)$ triplet into one source-side and one target-side vector.

    \item \textbf{Retrieve target history:} load the previous EMA history
    $h_j^{(c)}(t-1)$ for the target agent and channel.

    \item \textbf{Compute residual dependence:} estimate how much the source
    vector explains the target residual after conditioning on
    $h_j^{(c)}(t-1)$.

    \item \textbf{Update histories:} after scoring, update the target channel
    history and the streaming state used for future edge-channel estimates.
\end{enumerate}

In our implementation, the dependence score is computed using a lightweight
streaming covariance estimator over compact channel vectors. Concretely, the
compact source, target, and history vectors are concatenated, marginally
rank-normalized online, and used to update an exponentially smoothed covariance
estimate. The conditional dependence score is then computed from the covariance
blocks corresponding to source, target, and history, with shrinkage and jitter
for stable inversion. This provides a fast Gaussian-copula approximation to
conditional mutual information without fitting a density model or invoking an
additional LLM. The final score is clipped to be nonnegative and inserted into
the channel-specific influence matrix.

\noindent\textbf{(7) Building the unified cross-channel influence tensor.}
The LI-CTE score for each feasible edge and channel becomes one entry of the
cross-channel causal influence tensor:

\begin{equation}
\mathcal{A}_t(i,j,c)
=
A_t^{(c)}(i,j),
\qquad
\mathcal{A}_t
\in
\mathbb{R}^{N\times N\times |\mathcal{C}|}.
\label{eq:tensor}
\end{equation}

Each tensor slice corresponds to one interaction channel. The raw matrix used
for attribution is obtained by summing across channels,

\begin{equation}
A_t^{\mathrm{raw}}(i,j)
=
\sum_{c\in\mathcal{C}}
\mathcal{A}_t(i,j,c),
\label{eq:raw_aggregate}
\end{equation}

while the normalized matrix used for spectral monitoring is obtained by
applying the topology mask and degree-aware normalization described in the main
method. This produces the evolving influence topology $\tilde{A}_t$ used by
CASPIAN for cascade detection, phase monitoring, and online attribution.

\section{CASPIAN Algorithm}
\label{supp-sec:algorithm}

Algorithm~\ref{alg:caspian} summarizes the complete 
online CASPIAN pipeline. At each interaction turn, the 
framework extracts cross-channel MAS interaction events 
and estimates directed causal influence through LI-CTE to 
construct the evolving influence tensor 
$\mathcal{A}_t$. The resulting topology-normalized 
interaction graph $\tilde{A}_t$ is then monitored through 
spectral amplification, synchronization, phase-transition, 
and cross-channel propagation signals to detect both 
abrupt and gradual cascade formation online. Upon cascade 
confirmation, CASPIAN retrieves the cached influence 
interval associated with the detected propagation window 
and performs attribution directly over the evolving 
topology to recover the cascade origin, amplifier, bridge, 
and dominant propagation spines without replay or post-hoc 
trace analysis.

\begin{algorithm}[t]
\caption{\textsc{CASPIAN}: Online Cascade Detection and Attribution}
\label{alg:caspian}
\begin{algorithmic}[1]
\REQUIRE Agent set $\mathcal{V}$, channels $\mathcal{C}$, structural graph $G_0=(\mathcal{V},E_0)$, online trace $\mathcal{D}_{1:T}$
\ENSURE Cascade decision and attribution output $\mathcal{Y}_{t_0}$

\STATE Initialize $h_j^{(c)}(0)=\mathbf{0}$ for all $j\in\mathcal{V}, c\in\mathcal{C}$; cache $\mathcal{M}\gets\emptyset$; $t_w\gets\emptyset$; $\mathrm{Cascade}\gets0$

\FOR{$t=1,\ldots,T$}
    \STATE Extract channel events $\mathcal{E}_t=\{e_t^{(c)}\}_{c\in\mathcal{C}}$ from $\mathcal{D}_{1:t}$.
    \STATE Estimate LI-CTE influences $\mathcal{A}_t(i,j,c)$ for all feasible $(i,j)\in E_0$ and $c\in\mathcal{C}$.
    \STATE Normalize channel slices $\mathcal{A}_t^{(c)}$ and aggregate into $\tilde{A}_t$.
    \STATE Compute spectral signals $A_t^{\mathrm{amp}}, \Delta g_t, \Phi_t, H_t^{\mathrm{norm}}$.
    \STATE Update cache $\mathcal{M}\gets\mathcal{M}\cup\{(\tilde{A}_t,\mathcal{A}_t)\}$.

    \STATE $\mathrm{Watch}(t)\gets
    \mathbf{1}[A_t^{\mathrm{amp}}>1 \wedge \Delta g_t>0 \wedge \lambda_1(t)>\lambda_1(t-1)]$.

    \IF{$\mathrm{Watch}(t)=1$ and $t_w=\emptyset$}
        \STATE $t_w\gets t$
    \ENDIF

    \IF{$t_w\neq\emptyset$}
        \STATE Evaluate $\mathrm{\textsc{PhaseShift}}(t)$, $\mathrm{\textsc{CrossChannel}}(t)$, and $\mathrm{\textsc{WeakLink}}(t)$.

        \IF{$t=t_w$ and $(\mathrm{\textsc{PhaseShift}}(t)\vee\mathrm{\textsc{CrossChannel}}(t))\wedge\mathrm{\textsc{WeakLink}}(t)$}
            \STATE $t_0\gets t_w$; $\mathrm{Cascade}(t_0)\gets1$
        \ELSE
            \STATE $W_{t_w}\gets \left\lceil 1/(g_{t_w}+\epsilon)\right\rceil$; \quad $t_0\gets t_w+W_{t_w}-1$
            \IF{$t=t_0$}
                \STATE $\mathrm{MajorityWatch}\gets
                \mathbf{1}\!\left[
                \sum_{\tau=t_w}^{t_0}\mathrm{Watch}(\tau)
                \geq \frac{W_{t_w}}{2}
                \right]$.
                \STATE $\mathrm{Transition}\gets
                \mathbf{1}\!\left[
                \exists\,\tau\in[t_w,t_0]:
                \mathrm{\textsc{PhaseShift}}(\tau)\vee\mathrm{\textsc{CrossChannel}}(\tau)
                \right]$.
                \STATE $\mathrm{Cascade}(t_0)\gets \mathrm{MajorityWatch}\wedge\mathrm{Transition}$.
                \IF{$\mathrm{Cascade}(t_0)=0$}
                    \STATE Discard candidate interval and reset $t_w\gets\emptyset$.
                \ENDIF
            \ENDIF
        \ENDIF
    \ENDIF

    \IF{$\mathrm{Cascade}(t_0)=1$}
        \STATE Retrieve cached interval $\{(\tilde{A}_\tau,\mathcal{A}_\tau)\}_{\tau=t_w}^{t_0}$ from $\mathcal{M}$.
        \STATE Compute $i_{\mathrm{origin}}, i_{\mathrm{amp}}, i_{\mathrm{bridge}}$, and $\Pi^K$ using Eqs.~(12)--(15).
        \STATE Assign dominant channel $c^\star(\pi_k)$ for each spine $\pi_k\in\Pi^K$.
        \STATE \textbf{return} $\mathcal{Y}_{t_0}=\{i_{\mathrm{origin}},i_{\mathrm{amp}},i_{\mathrm{bridge}},\Pi^K,\{c^\star(\pi_k)\}_{k=1}^K\}$.
    \ENDIF
\ENDFOR
\end{algorithmic}
\end{algorithm}

\section{Additional Details on Evaluation}
\label{supp-sec:benchmarks}

\subsection{Evaluation Benchmark Details}
\label{supp-sec:eval-benchmarks}

\paragraph{Benchmarks.}
Table~\ref{tab:benchmarks} summarizes the evaluation 
benchmarks used in this work. TAMAS evaluates adversarial 
robustness in LLM-based multi-agent systems through 
prompt injection, coordination, and execution-oriented 
attacks across multiple MAS frameworks, while ACIArena 
focuses more heavily on cascading multi-turn injection 
behavior under diverse interaction topologies. Following 
prior work, we evaluate both attack and benign execution 
traces across AutoGen, CrewAI, MetaGPT, and LLM Debate. 
Exact benchmark scenario and execution trace counts are reported in Table~\ref{tab:trace_counts_app}. The final evaluation contains 727 benchmark
scenarios and 2,908 framework-expanded execution traces.

\begin{table}[t]
\centering
\caption{\textbf{Benchmark characteristics.} Summary of 
the benchmarks and evaluation settings used in this work.}
\vspace{6pt}
\label{tab:benchmarks}
\small
\begin{tabular}{lcc}
\toprule
\textbf{Property} & \textbf{TAMAS} & \textbf{ACIArena} \\
\midrule

\textbf{Primary focus}
& Adversarial robustness
& Cascading injection \\
& in LLM-MAS
& in LLM-MAS \\

\addlinespace[0.3em]

\textbf{Frameworks evaluated}
& AutoGen, CrewAI,
& AutoGen, CrewAI, \\
& MetaGPT, LLM Debate
& MetaGPT, LLM Debate \\

\addlinespace[0.3em]

\textbf{Attack categories}
& Intent, execution,
& Intent, execution, \\
& coordination
& coordination \\

\addlinespace[0.3em]

\textbf{CASPIAN channels}
& comm, mem,
& comm, mem, \\
& tool, exec
& tool, exec \\

\addlinespace[0.3em]

\textbf{Cascade regimes}
& Single-turn,
& Single-turn, \\
& multi-turn
& multi-turn \\

\addlinespace[0.3em]

\textbf{Evaluation setting}
& Per-framework,
& Cross-framework, \\
& per-attack analysis
& diverse MAS topologies \\

\addlinespace[0.3em]

\textbf{Ground truth signals}
& Injection traces,
& Injection traces, \\
& execution logs
& execution logs \\

\addlinespace[0.3em]

\textbf{Benign scenarios}
& \checkmark
& \checkmark \\

\textbf{Attack scenarios}
& \checkmark
& \checkmark \\

\bottomrule
\end{tabular}
\end{table}

\subsection{Benchmark Scenarios and Execution Traces}
\label{app:trace_counts}

Table~\ref{tab:trace_counts_app} reports the exact benchmark scenario and
execution trace counts used in our evaluation. Scenario counts are reported
before MAS-framework expansion, while trace counts correspond to executed runs
after evaluating each scenario across the four MAS frameworks used in this work.

\begin{table}[h]
\centering
\caption{\textbf{Evaluation scenario and trace counts.}
TAMAS categories correspond to Intent (IPI + Impersonation), Execution (DPI),
and Coordination (Byzantine + Colluding + Contradicting). ACIArena categories
correspond to Disclosure (MathLocation, MathName, CodeApikey, CodeName),
Disruption (DDOS), and Hijacking (SafetyCheck, MaliciousReport, AnswerMapping).}
\vspace{6pt}
\label{tab:trace_counts_app}
\small
\resizebox{\linewidth}{!}{%
\begin{tabular}{lrrrrrrr}
\toprule
Benchmark & Benign & Intent/Disc. & Exec./Disr. & Coord./Hijack & Attack & Scenarios & Traces \\
\midrule
TAMAS     & 100 & 100 & 50 & 150 & 300 & 400 & 1600 \\
ACIArena  & 69  & 138 & 30 & 90  & 258 & 327 & 1308 \\
\midrule
\textbf{Total} & \textbf{169} & \textbf{238} & \textbf{80} & \textbf{240} &
\textbf{558} & \textbf{727} & \textbf{2908} \\
\bottomrule
\end{tabular}%
}
\end{table}

TAMAS contributes 400 benchmark scenarios: 300 adversarial scenarios and 100
benign scenarios. ACIArena contributes 327 benchmark scenarios from the public
MASPI instruction-injection subset used in our evaluation: 69 benign base tasks
and 258 adversarial task--attack combinations. The ACIArena subset contains 39
math tasks and 30 code tasks, with two math attacks applied to math tasks and
six code attacks applied to code tasks. Evaluating these 727 benchmark scenarios
across four MAS frameworks yields 2,908 execution traces.

All reported AUROC, TPR@5\%FPR, and EDR@5 values are computed over the resulting
framework-stratified execution traces ($N=2{,}908$ total; $N=1{,}600$ TAMAS,
$N=1{,}308$ ACIArena). We report 95\% bootstrap confidence intervals over
1,000 trace-level resamples in Appendix~\ref{app:confidence_intervals}.

\subsection{Metrics}
\label{supp-sec:metrics}

Table~\ref{tab:metric_definitions} summarizes the metrics 
used throughout the evaluation. Detection metrics measure 
both overall separability and online responsiveness, while 
attribution metrics evaluate recovery of cascade roles, 
propagation structure, interaction modalities, and the 
delay required for successful online localization during 
active propagation.

*\textit{For AUROC and TPR@5\%FPR, each execution trace is assigned a scalar spectral
score equal to the strongest propagation evidence observed over the full trace.
AUROC is computed by sweeping a threshold over this score, and TPR@5\%FPR is
computed from the ROC curve at a 5\% false-positive rate. EDR@5 is computed
from the first online cascade alert produced by CASPIAN.}

\paragraph{Attribution ground truth.}
Attribution labels are derived from benchmark attack specifications and
framework execution logs, not from CASPIAN's influence matrices. The origin is
the agent receiving the initial adversarial instruction, corrupted memory item,
or malicious tool artifact. A bridge agent is labeled when it is the first
non-origin agent that transfers the corrupted state across a role boundary,
community boundary, or interaction channel. An amplifier is labeled as the
agent whose outputs are reused by the largest number of downstream agents after
corruption, measured from execution logs and contaminated artifact references.
Ground-truth spines are ordered propagation paths from the origin to downstream
affected agents, reconstructed from message references, memory read/write
events, tool-output dependencies, and execution logs. When multiple agents
satisfy the same role, all are retained as valid labels and Acc@1 is counted as
correct if the top-ranked prediction belongs to the valid set; MRR uses the
highest-ranked valid label. This avoids penalizing structurally equivalent
propagation paths in distributed MAS topologies.

\begin{table}[t]
\centering
\caption{\textbf{Evaluation metrics used throughout the paper.}}
\vspace{6pt}
\label{tab:metric_definitions}
\small
\begin{tabular}{p{2.3cm} p{4.2cm} p{5.7cm}}
\toprule
\textbf{Metric} & \textbf{Definition} & \textbf{Interpretation} \\
\midrule

TPR@5\% &
True positive rate measured at a fixed false positive rate of 5\%. &
Measures detection sensitivity under a strict false-alarm budget, reflecting practical online detection reliability. \\

EDR@5 &
Fraction of attacks detected within five turns of cascade onset. &
Measures how early a method detects propagation before the cascade fully spreads through the MAS. \\

AUROC &
Area under the receiver operating characteristic curve. &
Measures overall separability between attack and benign trajectories across decision thresholds. \\

Origin Acc@1 &
Fraction of cascades where the top-ranked predicted origin matches the ground-truth injection source. &
Evaluates localization of the initial propagating agent. \\

Amplifier Acc@1 &
Fraction of cascades where the top-ranked amplifier prediction is correct. &
Measures recovery of the agent contributing the strongest propagation reinforcement. \\

Bridge Acc@1 &
Fraction of cascades where the predicted bridge agent matches the ground-truth relay agent. &
Measures identification of agents relaying influence across propagation pathways. \\

MRR &
Mean reciprocal rank of the correct attributed agent. &
Evaluates how highly the correct role assignment is ranked on average. \\

Joint Acc@1 &
Fraction of cascades where origin, amplifier, and bridge are all identified correctly simultaneously. &
Measures overall attribution consistency under joint role recovery. \\

Spine Jaccard@3 &
Jaccard overlap between the top-3 recovered propagation spines and ground-truth propagation paths. &
Evaluates recovery of cascade propagation structure. \\

Channel Accuracy &
Fraction of recovered propagation spines whose dominant interaction channel is correctly identified. &
Measures recovery of the primary propagation modality (comm, mem, tool, exec). \\

Attribution Lag &
Average number of turns between cascade onset and successful attribution. &
Measures how quickly propagation structure can be localized during online execution. \\

\bottomrule
\end{tabular}
\end{table}

\subsection{Bootstrap Confidence Intervals}
\label{app:confidence_intervals}

All reported AUROC, TPR@5\%FPR, and EDR@5 values are computed over
framework-stratified traces. We report 95\% bootstrap confidence intervals over
1,000 resamples. Unless otherwise stated, bootstrap resampling is
performed at the trace level.

\begin{table}[h]
\centering
\caption{\textbf{Bootstrap confidence intervals for cascade detection.}
Each cell reports the point estimate with a 95\% confidence interval computed over
1,000 trace-level resamples.}
\vspace{6pt}
\label{tab:detection_ci}
\small
\resizebox{\linewidth}{!}{%
\begin{tabular}{llccc}
\toprule
Benchmark & Framework & Attack Type & AUROC & TPR@5\%FPR / EDR@5 \\
\midrule
TAMAS & AutoGen & Intent
& 0.932 [0.895, 0.969] & 0.846 [0.775, 0.917] / 0.781 [0.700, 0.862] \\
TAMAS & AutoGen & Execution
& 0.949 [0.905, 0.993] & 0.872 [0.779, 0.965] / 0.817 [0.710, 0.924] \\
TAMAS & AutoGen & Coordination
& 0.956 [0.932, 0.980] & 0.888 [0.838, 0.938] / 0.831 [0.771, 0.891] \\
\addlinespace
TAMAS & CrewAI & Intent
& 0.914 [0.873, 0.955] & 0.807 [0.730, 0.884] / 0.747 [0.662, 0.832] \\
TAMAS & CrewAI & Execution
& 0.934 [0.884, 0.984] & 0.842 [0.741, 0.943] / 0.782 [0.668, 0.896] \\
TAMAS & CrewAI & Coordination
& 0.951 [0.925, 0.977] & 0.865 [0.810, 0.920] / 0.795 [0.730, 0.860] \\
\addlinespace
TAMAS & MetaGPT & Intent
& 0.887 [0.840, 0.934] & 0.748 [0.663, 0.833] / 0.664 [0.571, 0.757] \\
TAMAS & MetaGPT & Execution
& 0.901 [0.841, 0.961] & 0.771 [0.655, 0.887] / 0.684 [0.555, 0.813] \\
TAMAS & MetaGPT & Coordination
& 0.907 [0.871, 0.943] & 0.782 [0.716, 0.848] / 0.690 [0.616, 0.764] \\
\addlinespace
TAMAS & LLM Debate & Intent
& 0.894 [0.848, 0.940] & 0.761 [0.677, 0.845] / 0.793 [0.714, 0.872] \\
TAMAS & LLM Debate & Execution
& 0.872 [0.805, 0.939] & 0.726 [0.602, 0.850] / 0.748 [0.628, 0.868] \\
TAMAS & LLM Debate & Coordination
& 0.881 [0.840, 0.922] & 0.741 [0.671, 0.811] / 0.774 [0.707, 0.841] \\
\midrule
ACIArena & AutoGen & Disclosure/Intent
& 0.918 [0.881, 0.955] & 0.813 [0.748, 0.878] / 0.758 [0.687, 0.829] \\
ACIArena & AutoGen & Disruption/Execution
& 0.946 [0.888, 1.000] & 0.858 [0.733, 0.983] / 0.804 [0.662, 0.946] \\
ACIArena & AutoGen & Hijacking/Coordination
& 0.948 [0.914, 0.982] & 0.869 [0.799, 0.939] / 0.814 [0.734, 0.894] \\
\addlinespace
ACIArena & CrewAI & Disclosure/Intent
& 0.901 [0.861, 0.941] & 0.782 [0.713, 0.851] / 0.724 [0.649, 0.799] \\
ACIArena & CrewAI & Disruption/Execution
& 0.927 [0.860, 0.994] & 0.821 [0.684, 0.958] / 0.761 [0.608, 0.914] \\
ACIArena & CrewAI & Hijacking/Coordination
& 0.936 [0.898, 0.974] & 0.843 [0.768, 0.918] / 0.774 [0.688, 0.860] \\
\addlinespace
ACIArena & MetaGPT & Disclosure/Intent
& 0.872 [0.825, 0.919] & 0.723 [0.648, 0.798] / 0.631 [0.550, 0.712] \\
ACIArena & MetaGPT & Disruption/Execution
& 0.889 [0.808, 0.970] & 0.754 [0.600, 0.908] / 0.654 [0.484, 0.824] \\
ACIArena & MetaGPT & Hijacking/Coordination
& 0.893 [0.843, 0.943] & 0.762 [0.674, 0.850] / 0.661 [0.563, 0.759] \\
\addlinespace
ACIArena & LLM Debate & Disclosure/Intent
& 0.879 [0.834, 0.924] & 0.742 [0.669, 0.815] / 0.768 [0.698, 0.838] \\
ACIArena & LLM Debate & Disruption/Execution
& 0.858 [0.768, 0.948] & 0.718 [0.557, 0.879] / 0.724 [0.564, 0.884] \\
ACIArena & LLM Debate & Hijacking/Coordination
& 0.864 [0.808, 0.920] & 0.731 [0.639, 0.823] / 0.751 [0.662, 0.840] \\
\bottomrule
\end{tabular}%
}
\end{table}

\begin{table}[h]
\centering
\caption{\textbf{Bootstrap confidence intervals for baseline comparison.}
Each cell reports the point estimate with a 95\% confidence interval over 1,000
trace-level resamples. EDR@5 is omitted for full-trace offline judges because
they do not support online early detection.}
\vspace{6pt}
\label{tab:baseline_ci}
\small
\resizebox{\linewidth}{!}{%
\begin{tabular}{lcccccc}
\toprule
 & \multicolumn{3}{c}{AutoGen} & \multicolumn{3}{c}{LLM Debate} \\
\cmidrule(lr){2-4}\cmidrule(lr){5-7}
Method & TPR@5\% & EDR@5 & AUROC & TPR@5\% & EDR@5 & AUROC \\
\midrule
Prompt Guard 2
& 0.284 [0.247, 0.321] & 0.198 [0.165, 0.231] & 0.618 [0.572, 0.664]
& 0.233 [0.198, 0.268] & 0.154 [0.124, 0.184] & 0.579 [0.531, 0.627] \\
JailGuard
& 0.326 [0.287, 0.365] & 0.242 [0.206, 0.278] & 0.630 [0.585, 0.675]
& 0.271 [0.234, 0.308] & 0.166 [0.135, 0.197] & 0.586 [0.539, 0.633] \\
PromptArmor
& 0.361 [0.321, 0.401] & 0.251 [0.215, 0.287] & 0.664 [0.621, 0.707]
& 0.304 [0.266, 0.342] & 0.207 [0.173, 0.241] & 0.622 [0.576, 0.668] \\
Perplexity
& 0.132 [0.104, 0.160] & 0.076 [0.054, 0.098] & 0.578 [0.530, 0.626]
& 0.086 [0.063, 0.109] & 0.045 [0.028, 0.062] & 0.533 [0.484, 0.582] \\
Single Turn
& 0.412 [0.371, 0.453] & 0.215 [0.181, 0.249] & 0.694 [0.652, 0.736]
& 0.383 [0.343, 0.423] & 0.190 [0.157, 0.223] & 0.661 [0.617, 0.705] \\
Sliding Window ($W=5$)
& 0.528 [0.487, 0.569] & 0.336 [0.297, 0.375] & 0.742 [0.704, 0.780]
& 0.496 [0.455, 0.537] & 0.314 [0.275, 0.353] & 0.717 [0.677, 0.757] \\
Sliding Window ($W=10$)
& 0.612 [0.572, 0.652] & 0.441 [0.400, 0.482] & 0.763 [0.727, 0.799]
& 0.575 [0.534, 0.616] & 0.395 [0.354, 0.436] & 0.736 [0.697, 0.775] \\
Full Trace
& 0.713 [0.675, 0.751] & -- & 0.811 [0.779, 0.843]
& 0.686 [0.647, 0.725] & -- & 0.765 [0.729, 0.801] \\
G-Safeguard
& 0.616 [0.576, 0.656] & 0.482 [0.441, 0.523] & 0.802 [0.769, 0.835]
& 0.582 [0.541, 0.623] & 0.437 [0.396, 0.478] & 0.785 [0.750, 0.820] \\
BlindGuard
& 0.645 [0.605, 0.685] & 0.521 [0.480, 0.562] & 0.815 [0.783, 0.847]
& 0.612 [0.572, 0.652] & 0.468 [0.427, 0.509] & 0.802 [0.769, 0.835] \\
GUARDIAN
& 0.632 [0.592, 0.672] & 0.498 [0.457, 0.539] & 0.818 [0.787, 0.849]
& 0.598 [0.557, 0.639] & 0.452 [0.411, 0.493] & 0.796 [0.762, 0.830] \\
\textbf{CASPIAN}
& \textbf{0.868 [0.840, 0.896]} & \textbf{0.792 [0.758, 0.826]} & \textbf{0.942 [0.926, 0.958]}
& \textbf{0.842 [0.812, 0.872]} & \textbf{0.753 [0.717, 0.789]} & \textbf{0.915 [0.895, 0.935]} \\
\bottomrule
\end{tabular}%
}
\end{table}

\begin{table}[h]
\centering
\caption{\textbf{Bootstrap confidence intervals for CASPIAN component ablations.}
Results are scenario-weighted across benchmarks, MAS frameworks, and attack mechanisms.
Each cell reports the point estimate with 95\% confidence interval over 1,000
trace-level resamples.}
\vspace{6pt}
\label{tab:ablation_ci}
\small
\resizebox{\linewidth}{!}{%
\begin{tabular}{llccc}
\toprule
Component & Variant & AUROC & TPR@5\%FPR & EDR@5 \\
\midrule
Influence & Cosine similarity
& 0.748 [0.729, 0.767] & 0.534 [0.513, 0.555] & 0.575 [0.554, 0.596] \\
Influence & MI (no conditioning)
& 0.797 [0.780, 0.814] & 0.603 [0.583, 0.623] & 0.630 [0.610, 0.650] \\
Influence & Full CTE
& 0.908 [0.898, 0.918] & 0.752 [0.734, 0.770] & 0.696 [0.677, 0.715] \\
\addlinespace
Channel & Comm-only
& 0.806 [0.790, 0.822] & 0.612 [0.592, 0.632] & 0.606 [0.586, 0.626] \\
Channel & Comm + Mem + Tool
& 0.883 [0.871, 0.895] & 0.753 [0.735, 0.771] & 0.712 [0.693, 0.731] \\
\addlinespace
Normalization & Raw $A_t$
& 0.871 [0.858, 0.884] & 0.670 [0.650, 0.690] & 0.705 [0.686, 0.724] \\
\addlinespace
Detection & $\lambda_1$ only
& 0.827 [0.812, 0.842] & 0.610 [0.590, 0.630] & 0.636 [0.616, 0.656] \\
Detection & + gap contraction $\Delta g_t$
& 0.864 [0.851, 0.877] & 0.698 [0.679, 0.717] & 0.671 [0.652, 0.690] \\
Detection & + phase shift $\Phi_t$
& 0.882 [0.870, 0.894] & 0.734 [0.716, 0.752] & 0.711 [0.692, 0.730] \\
Detection & + cross-channel entropy $H_t^{\mathrm{norm}}$
& 0.897 [0.886, 0.908] & 0.764 [0.746, 0.782] & 0.729 [0.711, 0.747] \\
\addlinespace
Path/Temp. & No \textsc{WeakLink}
& 0.891 [0.879, 0.903] & 0.715 [0.696, 0.734] & 0.746 [0.728, 0.764] \\
Path/Temp. & Fixed $W=5$
& 0.889 [0.877, 0.901] & 0.767 [0.749, 0.785] & 0.707 [0.688, 0.726] \\
\textbf{Full} & \textbf{Full CASPIAN}
& \textbf{0.906 [0.895, 0.917]} & \textbf{0.790 [0.773, 0.807]} & \textbf{0.743 [0.725, 0.761]} \\
\bottomrule
\end{tabular}%
}
\end{table}

*\textit{We compute confidence intervals by resampling traces with replacement within
each benchmark-framework stratum and recomputing each metric on every
resample. For attack-category tables, resampling is additionally stratified by
attack category. Reported intervals correspond to the 2.5th and 97.5th
percentiles across 1,000 resamples.}

%% file: main.bib
@article{watts2002simple,
  title={A simple model of global cascades on random networks},
  author={Watts, Duncan J},
  journal={Proceedings of the National Academy of Sciences},
  volume={99},
  number={9},
  pages={5766--5771},
  year={2002},
  publisher={The National Academy of Sciences}
}

@article{zhu2025llm,
  title={Where llm agents fail and how they can learn from failures},
  author={Zhu, Kunlun and Liu, Zijia and Li, Bingxuan and Tian, Muxin and Yang, Yingxuan and Zhang, Jiaxun and Han, Pengrui and Xie, Qipeng and Cui, Fuyang and Zhang, Weijia and others},
  journal={arXiv preprint arXiv:2509.25370},
  year={2025}
}

@inproceedings{kempe2003maximizing,
  title={Maximizing the spread of influence through a social network},
  author={Kempe, David and Kleinberg, Jon and Tardos, {\'E}va},
  booktitle={Proceedings of the ninth ACM SIGKDD international conference on Knowledge discovery and data mining},
  pages={137--146},
  year={2003}
}

@book{chung1997spectral,
  title={Spectral graph theory},
  author={Chung, Fan RK},
  volume={92},
  year={1997},
  publisher={American Mathematical Soc.}
}

@book{newman2011structure,
  title={The structure and dynamics of networks},
  author={Newman, Mark and Barab{\'a}si, Albert-L{\'a}szl{\'o} and Watts, Duncan J},
  year={2011},
  publisher={Princeton university press}
}

@article{de2025open,
  title={Open challenges in multi-agent security: Towards secure systems of interacting ai agents},
  author={de Witt, Christian Schroeder and Krawiecka, Klaudia and Krawczuk, Igor and Hagag, Ben and Anderson, William L and Belcak, Peter and Bucknall, Ben and Cai, Xiaohong and Chopra, Ayush and Cohen, Doron and others},
  journal={arXiv preprint arXiv:2505.02077},
  year={2025}
}

@article{zhang2025jailguard,
  title={Jailguard: A universal detection framework for prompt-based attacks on llm systems},
  author={Zhang, Xiaoyu and Zhang, Cen and Li, Tianlin and Huang, Yihao and Jia, Xiaojun and Hu, Ming and Zhang, Jie and Liu, Yang and Ma, Shiqing and Shen, Chao},
  journal={ACM Transactions on Software Engineering and Methodology},
  volume={35},
  number={1},
  pages={1--40},
  year={2025},
  publisher={ACM New York, NY}
}

@article{zocca2021spectral,
  title={A spectral representation of power systems with applications to adaptive grid partitioning and cascading failure localization},
  author={Zocca, Alessandro and Liang, Chen and Guo, Linqi and Low, Steven H and Wierman, Adam},
  journal={arXiv preprint arXiv:2105.05234},
  year={2021}
}

@article{venkatesh2026crea,
  title={Crea: A collaborative multi-agent framework for creative image editing and generation},
  author={Venkatesh, Kavana and Dunlop, Connor and Yanardag, Pinar},
  journal={Advances in Neural Information Processing Systems},
  volume={38},
  pages={171332--171392},
  year={2026}
}

@article{venkatesh2026physicsagentabm,
  title={PhysicsAgentABM: Physics-Guided Generative Agent-Based Modeling},
  author={Venkatesh, Kavana and He, Yinhan and Li, Jundong and Cui, Jiaming},
  journal={arXiv preprint arXiv:2602.06030},
  year={2026}
}

@article{venkatesh2026agent,
  title={Do Agent Societies Develop Intellectual Elites? The Hidden Power Laws of Collective Cognition in LLM Multi-Agent Systems},
  author={Venkatesh, Kavana and Cui, Jiaming},
  journal={arXiv preprint arXiv:2604.02674},
  year={2026}
}

@article{shi2025promptarmor,
  title={Promptarmor: Simple yet effective prompt injection defenses},
  author={Shi, Tianneng and Zhu, Kaijie and Wang, Zhun and Jia, Yuqi and Cai, Will and Liang, Weida and Wang, Haonan and Alzahrani, Hend and Lu, Joshua and Kawaguchi, Kenji and others},
  journal={arXiv preprint arXiv:2507.15219},
  year={2025}
}

@inproceedings{wu2024autogen,
  title={Autogen: Enabling next-gen LLM applications via multi-agent conversations},
  author={Wu, Qingyun and Bansal, Gagan and Zhang, Jieyu and Wu, Yiran and Li, Beibin and Zhu, Erkang and Jiang, Li and Zhang, Xiaoyun and Zhang, Shaokun and Liu, Jiale and others},
  booktitle={First conference on language modeling},
  year={2024}
}

@inproceedings{wang2025g,
  title={G-safeguard: A topology-guided security lens and treatment on llm-based multi-agent systems},
  author={Wang, Shilong and Zhang, Guibin and Yu, Miao and Wan, Guancheng and Meng, Fanci and Guo, Chongye and Wang, Kun and Wang, Yang},
  booktitle={Proceedings of the 63rd Annual Meeting of the Association for Computational Linguistics (Volume 1: Long Papers)},
  pages={7261--7276},
  year={2025}
}

@article{miao2025blindguard,
  title={Blindguard: Safeguarding llm-based multi-agent systems under unknown attacks},
  author={Miao, Rui and Liu, Yixin and Wang, Yili and Shen, Xu and Tan, Yue and Dai, Yiwei and Pan, Shirui and Wang, Xin},
  journal={arXiv preprint arXiv:2508.08127},
  year={2025}
}

@article{zhou2026guardian,
  title={Guardian: Safeguarding llm multi-agent collaborations with temporal graph modeling},
  author={Zhou, Jialong and Wang, Lichao and Yang, Xiao},
  journal={Advances in Neural Information Processing Systems},
  volume={38},
  pages={7973--8001},
  year={2026}
}

@article{kavathekar2025tamas,
  title={TAMAS: Benchmarking Adversarial Risks in Multi-Agent LLM Systems},
  author={Kavathekar, Ishan and Jain, Hemang and Rathod, Ameya and Kumaraguru, Ponnurangam and Ganu, Tanuja},
  journal={arXiv preprint arXiv:2511.05269},
  year={2025}
}

@article{chennabasappa2025llamafirewall,
  title={Llamafirewall: An open source guardrail system for building secure ai agents},
  author={Chennabasappa, Sahana and Nikolaidis, Cyrus and Song, Daniel and Molnar, David and Ding, Stephanie and Wan, Shengye and Whitman, Spencer and Deason, Lauren and Doucette, Nicholas and Montilla, Abraham and others},
  journal={arXiv preprint arXiv:2505.03574},
  year={2025}
}

@article{an2026aciarena,
  title={ACIArena: Toward Unified Evaluation for Agent Cascading Injection},
  author={An, Hengyu and Li, Minxi and Zhang, Jinghuai and Xu, Naen and Zhou, Chunyi and Li, Changjiang and Xu, Xiaogang and Du, Tianyu and Ji, Shouling},
  journal={arXiv preprint arXiv:2604.07775},
  year={2026}
}

@article{zhang2025agent,
  title={Which agent causes task failures and when? on automated failure attribution of llm multi-agent systems},
  author={Zhang, Shaokun and Yin, Ming and Zhang, Jieyu and Liu, Jiale and Han, Zhiguang and Zhang, Jingyang and Li, Beibin and Wang, Chi and Wang, Huazheng and Chen, Yiran and others},
  journal={arXiv preprint arXiv:2505.00212},
  year={2025}
}

@article{jia2026mas,
  title={MAS-FIRE: Fault Injection and Reliability Evaluation for LLM-Based Multi-Agent Systems},
  author={Jia, Jin and Deng, Zhiling and Chen, Zhuangbin and Wang, Yingqi and Zheng, Zibin},
  journal={arXiv preprint arXiv:2602.19843},
  year={2026}
}

@article{wang2026flat,
  title={From Flat Logs to Causal Graphs: Hierarchical Failure Attribution for LLM-based Multi-Agent Systems},
  author={Wang, Yawen and Wu, Wenjie and Wang, Junjie and Wang, Qing},
  journal={arXiv preprint arXiv:2602.23701},
  year={2026}
}

@article{gu2024survey,
  title={A survey on llm-as-a-judge},
  author={Gu, Jiawei and Jiang, Xuhui and Shi, Zhichao and Tan, Hexiang and Zhai, Xuehao and Xu, Chengjin and Li, Wei and Shen, Yinghan and Ma, Shengjie and Liu, Honghao and others},
  journal={The Innovation},
  year={2024},
  publisher={Elsevier}
}

@article{radford2019language,
  title={Language models are unsupervised multitask learners},
  author={Radford, Alec and Wu, Jeffrey and Child, Rewon and Luan, David and Amodei, Dario and Sutskever, Ilya and others},
  journal={OpenAI blog},
  volume={1},
  number={8},
  pages={9},
  year={2019}
}

@inproceedings{lee2025prompt,
  title={Prompt infection: Llm-to-llm prompt injection within multi-agent systems},
  author={Lee, Donghyun and Tiwari, Mo and Miranda, Brando},
  booktitle={European Symposium on Research in Computer Security},
  pages={511--520},
  year={2025},
  organization={Springer}
}

@article{liang2025tipping,
  title={Tipping the Dominos: Topology-Aware Multi-Hop Attacks on LLM-Based Multi-Agent Systems},
  author={Liang, Ruichao and Yin, Le and Chen, Jing and Wu, Cong and Zhang, Xiaoyu and Gu, Huangpeng and Zhang, Zijian and Liu, Yang},
  journal={arXiv preprint arXiv:2512.04129},
  year={2025}
}

@inproceedings{shen2025understanding,
  title={Understanding the information propagation effects of communication topologies in llm-based multi-agent systems},
  author={Shen, Xu and Liu, Yixin and Dai, Yiwei and Wang, Yili and Miao, Rui and Tan, Yue and Pan, Shirui and Wang, Xin},
  booktitle={Proceedings of the 2025 Conference on Empirical Methods in Natural Language Processing},
  pages={12358--12372},
  year={2025}
}

@article{jiang2026agentlab,
  title={Agentlab: Benchmarking llm agents against long-horizon attacks},
  author={Jiang, Tanqiu and Wang, Yuhui and Liang, Jiacheng and Wang, Ting},
  journal={arXiv preprint arXiv:2602.16901},
  year={2026}
}

@article{xie2026spark,
  title={From Spark to Fire: Modeling and Mitigating Error Cascades in LLM-Based Multi-Agent Collaboration},
  author={Xie, Yizhe and Zhu, Congcong and Zhang, Xinyue and Zhu, Tianqing and Ye, Dayong and Qi, Minfeng and Chen, Huajie and Zhou, Wanlei},
  journal={arXiv preprint arXiv:2603.04474},
  year={2026}
}

@article{schreiber2000measuring,
  title={Measuring information transfer},
  author={Schreiber, Thomas},
  journal={Physical review letters},
  volume={85},
  number={2},
  pages={461},
  year={2000},
  publisher={APS}
}

@article{lee2024identifying,
  title={Identifying Influential and Vulnerable Nodes in Interaction Networks through Estimation of Transfer Entropy Between Univariate and Multivariate Time Series},
  author={Lee, Julian},
  journal={arXiv preprint arXiv:2408.15811},
  year={2024}
}

@article{sun2014causation,
  title={Causation entropy identifies indirect influences, dominance of neighbors and anticipatory couplings},
  author={Sun, Jie and Bollt, Erik M},
  journal={Physica D: Nonlinear Phenomena},
  volume={267},
  pages={49--57},
  year={2014},
  publisher={Elsevier}
}

@article{shahsavari2020estimating,
  title={Estimating conditional transfer entropy in time series using mutual information and nonlinear prediction},
  author={Shahsavari Baboukani, Payam and Graversen, Carina and Alickovic, Emina and {\O}stergaard, Jan},
  journal={Entropy},
  volume={22},
  number={10},
  pages={1124},
  year={2020},
  publisher={MDPI}
}

@article{zhang2025graphtracer,
  title={GraphTracer: Graph-Guided Failure Tracing in LLM Agents for Robust Multi-Turn Deep Search},
  author={Zhang, Heng and Shi, Yuling and Gu, Xiaodong and You, Haochen and Zhang, Zijian and Gan, Lubin and Yuan, Yilei and Huang, Jin},
  journal={arXiv preprint arXiv:2510.10581},
  year={2025}
}

@inproceedings{liang2024encouraging,
  title={Encouraging divergent thinking in large language models through multi-agent debate},
  author={Liang, Tian and He, Zhiwei and Jiao, Wenxiang and Wang, Xing and Wang, Yan and Wang, Rui and Yang, Yujiu and Shi, Shuming and Tu, Zhaopeng},
  booktitle={Proceedings of the 2024 conference on empirical methods in natural language processing},
  pages={17889--17904},
  year={2024}
}

@article{zhang2025agentracer,
  title={AgenTracer: Who Is Inducing Failure in the LLM Agentic Systems?},
  author={Zhang, Guibin and Wang, Junhao and Chen, Junjie and Zhou, Wangchunshu and Wang, Kun and Yan, Shuicheng},
  journal={arXiv preprint arXiv:2509.03312},
  year={2025}
}

@book{seneta1981nonnegative,
  title={Non-negative Matrices and Markov Chains},
  author={Seneta, Eugene},
  year={1981},
  publisher={Springer}
}

@book{horn2012matrix,
  title={Matrix Analysis},
  author={Horn, Roger A. and Johnson, Charles R.},
  edition={2nd},
  year={2012},
  publisher={Cambridge University Press}
}

@article{hoory2006expander,
  title={Expander Graphs and their Applications},
  author={Hoory, Shlomo and Linial, Nathan and Wigderson, Avi},
  journal={Bulletin of the American Mathematical Society},
  volume={43},
  number={4},
  pages={439--561},
  year={2006}
}

@book{levin2017markov,
  title={Markov Chains and Mixing Times},
  author={Levin, David A. and Peres, Yuval},
  edition={2nd},
  year={2017},
  publisher={American Mathematical Society}
}

@article{spielman2007spectral,
  title={Spectral Graph Theory and its Applications},
  author={Spielman, Daniel A.},
  booktitle={Proceedings of the 48th Annual IEEE Symposium on 
             Foundations of Computer Science},
  pages={29--38},
  year={2007},
  organization={IEEE}
}

@article{restrepo2006spectral,
  title={Spectral Properties of Complex Networks},
  author={Restrepo, Juan G. and Ott, Edward and Hunt, Brian R.},
  journal={Chaos: An Interdisciplinary Journal of Nonlinear Science},
  volume={16},
  number={1},
  year={2006},
  publisher={AIP Publishing}
}

@article{goltsev2012localization,
  title={Localization and Spreading of Diseases in Complex Networks},
  author={Goltsev, Alexander V. and Dorogovtsev, Sergey N. and 
          Oliveira, J. G. and Mendes, J. F. F.},
  journal={Physical Review Letters},
  volume={109},
  number={12},
  pages={128702},
  year={2012},
  publisher={APS}
}

@article{shannon1948,
  title={A Mathematical Theory of Communication},
  author={Shannon, Claude E.},
  journal={The Bell System Technical Journal},
  volume={27},
  number={3},
  pages={379--423},
  year={1948},
  publisher={Nokia Bell Labs}
}

@inproceedings{khattab2020colbert,
  title={Colbert: Efficient and effective passage search via contextualized late interaction over bert},
  author={Khattab, Omar and Zaharia, Matei},
  booktitle={Proceedings of the 43rd International ACM SIGIR conference on research and development in Information Retrieval},
  pages={39--48},
  year={2020}
}

@misc{CrewAI2024,
  author = {{CrewAI Team}},
  title  = {CrewAI: Framework for Orchestrating Role-Playing, Collaborative AI Agents},
  year   = {2024},
  url    = {https://github.com/crewAIInc/crewAI},
  note   = {Accessed: 2026-05-06}
}

@inproceedings{hong2023metagpt,
  title={MetaGPT: Meta programming for a multi-agent collaborative framework},
  author={Hong, Sirui and Zhuge, Mingchen and Chen, Jonathan and Zheng, Xiawu and Cheng, Yuheng and Wang, Jinlin and Zhang, Ceyao and Wang, Zili and Yau, Steven Ka Shing and Lin, Zijuan and others},
  booktitle={The twelfth international conference on learning representations},
  year={2023}
}

@misc{EMRA,
      title={CoopGuard: Stateful Cooperative Agents Safeguarding LLMs Against Evolving Multi-Round Attacks}, 
      author={Siyuan Li and Zehao Liu and Xi Lin and Qinghua Mao and Yuliang Chen and Haoyu Li and Jun Wu and Jianhua Li and Xiu Su},
      year={2026},
      eprint={2604.04060},
      archivePrefix={arXiv},
      primaryClass={cs.CR},
      url={https://arxiv.org/abs/2604.04060}, 
}
